\documentclass[twocolumn,preprintnumbers,prd,superscriptaddress,nofootinbib,floatfix,showpacs,showkeys]{revtex4-1}
\usepackage{amsmath}
\usepackage{amsfonts}
\usepackage{amssymb}
\usepackage{graphicx}
\usepackage{color}
\usepackage{hyperref}
\usepackage{booktabs}
\usepackage{hyperref}
\usepackage{cleveref}
\usepackage{slashed}
\usepackage{braket}
\usepackage{mathtools}
\usepackage[rightcaption]{sidecap}
\usepackage{subfigure}
\usepackage{soul}
\usepackage{enumitem}
\usepackage{comment}
\usepackage{multirow}

\begin{document}

\title{The \emph{macroscopic precession model}: describing quasi-periodic oscillations including internal structures of test bodies}

\author{Gabriele Bianchini}
\email{gabriele.bianchini@studenti.unicam.it}
\affiliation{Universit\`a di Camerino, Via Madonna delle Carceri, Camerino, 62032, Italy.}

\author{Orlando Luongo}
\email{orlando.luongo@unicam.it}
\affiliation{Universit\`a di Camerino, Via Madonna delle Carceri, Camerino, 62032, Italy.}
\affiliation{INAF - Osservatorio Astronomico di Brera, Milano, Italy.}
\affiliation{Istituto Nazionale di Fisica Nucleare, Sezione di Perugia, Perugia, 06123, Italy.}
\affiliation{SUNY Polytechnic Institute, 13502 Utica, New York, USA.}
\affiliation{Al-Farabi Kazakh National University, Al-Farabi av. 71, 050040 Almaty, Kazakhstan.}

\author{Marco Muccino}
\email{marco.muccino@unicam.it}
\affiliation{Universit\`a di Camerino, Via Madonna delle Carceri, Camerino, 62032, Italy.}
\affiliation{Al-Farabi Kazakh National University, Al-Farabi av. 71, 050040 Almaty, Kazakhstan.}
\affiliation{Institute of Nuclear Physics, Ibragimova, 1, 050032 Almaty, Kazakhstan.}
\affiliation{ICRANet, Piazza della Repubblica 10, Pescara, 65122, Italy.}

\begin{abstract}
The relativistic precession model (RPM) is widely-considered as a benchmark framework to interpret quasi-periodic oscillations (QPOs), albeit several observational inconsistencies suggest that the model remains incomplete. The RPM ensures \emph{structureless test particles} and attributes precession to geodesic motion alone. Here, we refine the RPM by incorporating the internal structure of rotating test bodies, while preserving the test particle approximation (TPA), and propose a \emph{macroscopic precession model} (MPM) by means of the Mathisson-Papapetrou-Dixon (MPD) equations, applied to a Schwarzschild background, which introduces 1) a shift in the Keplerian frequency and 2) an \emph{effective spin correction} to the radial epicyclic frequency that, once the spin tensor is modeled, reproduces a quasi-Schwarzschild-de Sitter (SdS) correction. We apply the MPM to eight neutron star low mass X-ray binaries (NS-LMXBs), performing Markov chain Monte Carlo (MCMC) fits to twin kHz QPOs and find observational and statistical evidence in favor of precise power law spin reconstructions. Further, our model accurately predicts the $3:2$ frequency clustering, the disk boundaries and the NS masses. From the MPM model, we thus conclude that complexity of QPOs can be fully-described including the test particle internal structure.
\end{abstract}


\maketitle

\section{Introduction}\label{sec:intro}

QPOs are narrow, nearly stable but slowly drifting peaks observed in X-ray power spectra of accreting compact objects \cite{2004PASJ...56..553R,2004ApJ...609L..63A,torok05,2012ApJ...760..138T}, like LMXBs hosting black holes or NSs \cite{Lamb2008}, and active galactic nuclei on much larger scales \cite{mchardy06}. 

Their physical origin is still debated and no single model has reached universal consensus \cite{ingram19}. Nonetheless, QPOs are widely recognized as precision probes of the strong-field regime and as tools to constrain masses and spins of compact objects \cite{motta14}. However, this places an inconsistency: if their origin is not clearly known, how much QPOs may turn to be predictive?

In NS-LMXBs, QPO phenomenology depends on the source spectral state and position in the color-color diagram \cite{vanderklis06,swank04,vanderklis97}. Two main branches exist: Z sources, characterized by high accretion rates \cite{ingram19}, and Atoll sources, typically at lower rates. At higher frequencies, both classes display kHz QPOs, often showing twin peaks with centroid frequencies $\sim 0.3-1.2$~kHz, interpreted as the signature of orbital motion in the innermost accretion flow.

The RPM has become a reference paradigm for interpreting such signals \cite{belloni14,wang16}. It associates the observed frequencies with the azimuthal, radial, and vertical epicyclic modes of a test particle orbiting in a prescribed spacetime \cite{1999PhRvL..82...17S}. The analytic expressions depend on the assumed geometry, and the model has been used, for example, to estimate black hole spins in agreement with independent measurements \cite{motta14}. 

However, recent analyses  highlighted that the RPM tends to prefer regular or Schwarzschild-de~Sitter (SdS) metrics over standard Schwarzschild or Reissner-Nordström geometries \cite{boshkayev23a,boshkayev23b}. The overall finding  disfavors metric functions that show behaviors falling off faster than $r^{-1}$, and remains challenged by the observed $3:2$ frequency clustering \cite{lewin88}. 
As an immediate step forward, even the inclusion of anharmonic or quadrupolar corrections appear unable to fix the problem \cite{giambo25,Boshkayev:2023esw,torok05}, leaving unexpectedly the SdS solution as favored, without giving a clear explanation about the nature of the associated cosmological constant, $R_0$. Phrasing it differently, how is it possible that the standard geometry offered by Schwarzschild, or corrected by electric charge, like in the Reisser-Nordstrom, are unable to well reproduce QPO data, while a cosmological constant does?

In this work, we show that a severe RPM limitation is treating the accreting matter as a collection of structureless test particles, albeit real disks possess macroscopic angular momentum and internal structure.

Hence, we address this gap by \emph{introducing a MPM paradigm based on the MPD equations, keeping the TPA but endowing each effective ``particle'' with a classical intrinsic spin}. In our picture, the effective spin describes the internal rotation of an extended body and its coupling to curvature, generating \emph{de facto} off-diagonal contributions to the energy-momentum tensor, resulting into a spin-curvature coupling that modifies the radial epicyclic frequencies, once the geometry is fixed. For the sake of simplicity, we limit on the Schwarzschild background and show that a SdS-like correction is found, contributing to the radial and azimuthal frequencies with corrections that: 1) depend on the spin form, 2) are consequence of the underlying gauge adopted, 3) provide the same leading terms with departures that dominate at very large radii. To confront our model with data, we focus on eight NS-LMXBs with twin kHz QPOs: the Z sources Cir~X-1, GX~5-1, GX~17+2, GX~340+0, Sco~X-1 \cite{hasinger89,shirey98} and the Atoll sources 4U~1608-52, 4U~1728-34, and 4U~0614+091 \cite{hasinger89,vanstraaten00}, whose  lower and upper frequencies, viz. $f_{\rm L}$ and $f_{\rm U}$, span inside $\sim50$-$900\,$Hz and up to $\sim1200\,$Hz, respectively. Afterwards, we perform MCMC fits to the $f_{\rm L}$--$f_{\rm U}$ frequencies of the eight NS-LMXBs mentioned above, adopting a non-vanishing spin parametrized by power-law terms $\propto r^n$ with tightly constrained macroscopic index $n=2,3$. We thus show that the MPM introduces net improvements w.r.t. the pure Schwarzschild RPM framework and provides fits quite competitive with those of the RPM based on the SdS metric. As a consequence of our recipe, we simultaneously satisfy the TPA, reconstructing the radial extent of the disks. Our findings certify the need of internal structure for the test particles inside the disk, while justifying viable constraints highly-compatible with previous literature \cite{smith21,done07,klis89,tagger07}.

The paper is outlined as follows. In Sec.~\ref{MPM}, we present the MPM, starting from the RPM baseline and then introducing the MPD extension on a Schwarzschild background. In Sec.~\ref{sec:statistical},  we report our  statistical analysis, while in Sec.~\ref{sec:conclusions},  we summarize the main results and outline future developments.


\section{Introducing the macroscopic precession model}
\label{MPM}

We here construct the MPM, including into the RPM the presence of the spin of the orbiting test bodies. 

The dynamics of a structureless test particle of mass $m$ and four-velocity $\dot{x}^\mu=dx^\mu/d\tau$ is described by a Lagrangian $\mathbb{L}=m\,g_{\mu\nu}\dot{x}^\mu\dot{x}^\nu/2$, where $g_{\mu\nu}$ is the metric tensor that, for a static and spherical spacetime, provides specific energy $\varepsilon=-g_{tt} u^t$ and angular momentum $\ell = g_{\phi\phi} u^\phi$ as conserved quantities \cite{turimov22}.
The normalization of the four-velocity, i.e., $\dot{x}^\mu \dot{x}_\mu=-1$, provides $g_{rr}\dot{r}^2 + g_{\theta\theta}\dot{\theta}^2 + \mathcal V(r,\theta) = 0$. Stable circular orbits $x_0=(r_0,\theta_0)$ occur at the minima of the potential $\mathcal V$, where $\mathcal V(x_0)=\partial_r\mathcal V|_{x_0}=\partial_\theta\mathcal V|_{x_0}=0$.
For equatorial orbits, $\theta_0=\pi/2$, $\dot{r}=\dot{\theta}=0$, one finds the azimuthal and the radial angular frequencies \cite{bambi16}, respectively,
\begin{equation}
\label{Ophi_Or_RPM}
\Omega_\phi^2 = -\frac{\partial_r g_{tt}}{\partial_r g_{\phi\phi}}\qquad,\qquad
\Omega_r^2 = \frac{\partial_r^2\mathcal V|_{x_0}}{2g_{rr}(u^t)^2}\,,
\end{equation}
demanding small perturbations $r\to r_0+\delta r$ and $\theta\to\theta_0+\delta\theta$ to derive $\delta r'' + \Omega_r^2 \delta r = 0$, with prime denoting time derivatives. With these definitions, the RPM identifies the lower and upper QPOs with the periastron and Keplerian frequencies, respectively,
\begin{equation}
\label{QPO_freqs}
f_L=\frac{1}{2\pi}(\Omega_\phi-\Omega_r)\,,\qquad
f_U=\frac{\Omega_\phi}{2\pi}\,.
\end{equation}
These frequencies can be modified by accounting for the motion of a massive, spinning test body in a curved spacetime described by the MPD equations \cite{Poisson:2011nh,loomis17,Costa:2017kdr}
\begin{subequations}
\label{eq1}
\begin{align}
\frac{Dp^\mu}{D\tau} &= -\frac{1}{2}u^\pi S^{\rho\sigma}R^\mu_{\ \pi\rho\sigma}\,, \label{eq1a}\\
\frac{D S^{\mu\nu}}{D\tau} &= p^\mu u^\nu - p^\nu u^\mu\,, \label{eq1b}
\end{align}
\end{subequations}
where $S^{\mu\nu}$ is the antisymmetric spin tensor, $u^\mu=dx^\mu/d\tau$ is the kinematic four-velocity, $p^\mu$ the generalized four-momentum, $R^\lambda_{\ \pi\rho\sigma}$ the Riemann tensor, and $D/D\tau$ the covariant derivative along the worldline. The generalized four-momentum is now shifted by 
\begin{equation}
\label{eq3}
p^\mu = m u^\mu + u_\lambda \frac{D S^{\mu\lambda}}{D\tau}\,.
\end{equation}

To close the system, we impose the Tulczyjew-Dixon spin supplementary gauge condition \cite{Dixon:1979}
\begin{equation}
\label{eq2}
S^{\lambda\nu}p_\nu=0\,,
\end{equation}
and, in the presence of a Killing vector $\xi^\mu$, the MPD equations admit a conserved quantity
\begin{equation}
\label{eq4}
C[\xi]=p^\mu\xi_\mu+\frac{1}{2}\nabla_\mu\xi_\nu S^{\nu\mu}\,,
\end{equation}
where $\nabla_\mu$ denotes the covariant derivative. 
The TPA is preserved, as well as in the RPM, if one requires 
\begin{equation}
\label{eq20}
\kappa\equiv |S_0|(mr)^{-1}\ll1\,,
\end{equation}
with $\kappa$ a dimensionless parameter and $S_0^2=S_{\mu\nu}S^{\mu\nu}/2$, i.e., the intrinsic spin should be weak if compared to the curvature scale, to avoid significant particle backreaction on the background geometry\footnote{This may induce additional dynamical effects, e.g., 
self-forces and non-local history-dependent forces, see Refs.~\cite{Mino:1996nk,Quinn:2000wa,PoissonPoundVega2011,QuinnWald1997,DetweilerWhiting2003}. } \cite{bini17}.

\paragraph*{{{\bf \emph{MPM on a Schwarzschild background.}}}}
We now apply the MPD formalism to a Schwarzschild metric, where time translation and axial symmetry imply the existence of Killing vectors $\xi^\mu_{(t)}=\partial_t^\mu$ and $\xi^\mu_{(\phi)}=\partial_\phi^\mu$. Hence, assuming $dm/d\tau=0$, Eq.~\eqref{eq4} yields the generalized specific energy $\mathcal E$ and angular momentum $\mathcal L$,
\begin{subequations}
\label{eq5}
\begin{align}
-\mathcal E &= g_{tt}u^t + g_{tt} u_\lambda\frac{D\mathcal S^{t\lambda}}{D\tau} + \frac{1}{4}\mathcal S^{tr}\partial_rg_{tt}\,, \label{eq5a}\\
\mathcal L &= g_{\phi\phi}u^\phi + g_{\phi\phi}u_\lambda\frac{D\mathcal S^{\phi\lambda}}{D\tau} + \frac{1}{4}\mathcal S^{\phi r}\partial_rg_{\phi\phi}\,, \label{eq5b}
\end{align}
\end{subequations}
where $\mathcal S^{\mu\nu}=S^{\mu\nu}/m$ is the specific spin tensor. In the spinless limit, $\mathcal S^{\mu\nu}=0$, one recovers $\mathcal E=\varepsilon$ and $\mathcal L=\ell$.

Introducing the dynamical four-velocity $v^\mu=p^\mu/m$, the conserved quantities can be written as
\begin{subequations}
\label{eq7}
\begin{align}
\mathcal E + \Delta\mathcal E &= -g_{tt}v^t\,, 
\qquad \Delta\mathcal E = \frac{1}{4}\mathcal S^{tr}\partial_rg_{tt}\,, \label{eq7a}\\
\mathcal L - \Delta\mathcal L &= g_{\phi\phi}v^\phi\,,
\qquad \Delta\mathcal L = \frac{1}{4}\mathcal S^{\phi r}\partial_rg_{\phi\phi}\,. \label{eq7b}
\end{align}
\end{subequations}

For equatorial circular orbits we have $v^r=v^\theta=0$ and
\begin{equation}
\label{eq8}
v^t = \sqrt{-(g_{tt}+\Omega_\phi^2 g_{\phi\phi})^{-1}}\quad,\quad v^\phi = \Omega_\phi v^t\,.
\end{equation}
Now $\Omega_\phi=v^\phi/v^t$ generalizes the Keplerian frequency. Using Eqs.~\eqref{eq7}--\eqref{eq8}, the constants of motion become
\begin{subequations}
\label{eq9}
\begin{align}
\mathcal E + \Delta\mathcal E &= -g_{tt} \sqrt{-(g_{tt}+\Omega_\phi^2g_{\phi\phi})^{-1}}\,, \label{eq9a}\\
\mathcal L - \Delta\mathcal L &= g_{\phi\phi} \Omega_\phi \sqrt{-(g_{tt}+\Omega_\phi^2g_{\phi\phi})^{-1}}\,, \label{eq9b}
\end{align}
\end{subequations}
and the effective potential reads
\begin{equation}
\label{eqVbis}
\mathcal V(r,\theta) = 1 + g_{tt}^{-1}(\mathcal E+\Delta\mathcal E)^2 + g_{\phi\phi}^{-1}(\mathcal L-\Delta\mathcal L)^2\,.
\end{equation}
From $\partial_r\mathcal V|_{x_0}=0$, the modified $\Omega_\phi$ is obtained from 
\begin{equation}
\label{eq_for_Of}
v^t\left(\partial_rg_{tt}+\Omega_\phi^2\partial_rg_{\phi\phi}\right)
+2\left(\partial_r\Delta\mathcal E+\Omega_\phi\partial_r\Delta\mathcal L\right)=0\,,
\end{equation}

The spin structure is based on  $\mathcal S^{tr}$ and $\mathcal S^{\phi r}$ only in view of Eqs.~\eqref{eq2} and \eqref{eq7}, 
\begin{equation}
\label{eqS}
\mathcal S^{\mu\nu}=
\begin{pmatrix}
0 & \mathcal S^{tr} & 0 & 0\\
-\mathcal S^{tr} & 0 & 0 & \mathcal S^{r\phi}\\
0 & 0 & 0 & 0\\
0 & -\mathcal S^{r\phi} & 0 & 0
\end{pmatrix},
\end{equation}
ending up with
\begin{equation}
\label{eqSbis}
\mathcal S^{\phi r} = -\frac{g_{tt}}{g_{\phi\phi}}\frac{\mathcal S^{tr}}{\Omega_\phi}
= \left(1-\frac{2M}{r}\right)\frac{\mathcal S^{tr}}{r^2\Omega_\phi}\,,
\end{equation}
that furnishes
\begin{equation}
\label{eqdeltas}
\Delta\mathcal E = -\frac{M\mathcal S^{tr}}{2r^2}\,,\qquad
\Delta\mathcal L = \left(1-\frac{2M}{r}\right)\frac{\mathcal S^{tr}}{2r\Omega_\phi}\,.
\end{equation}

\begin{figure*}
\includegraphics[width=0.95\hsize,clip]{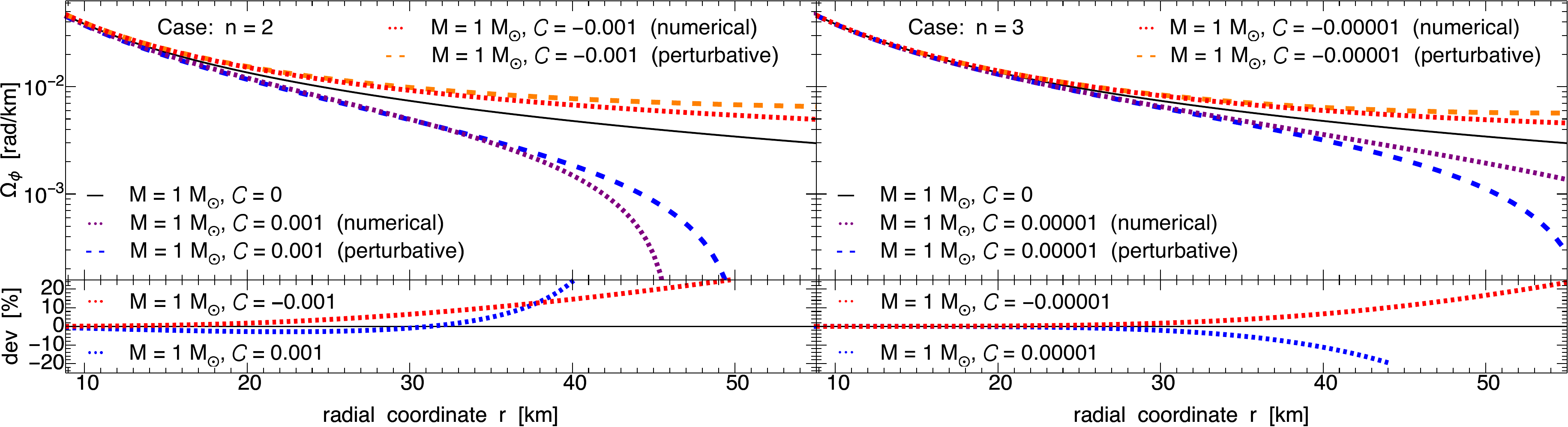}
\caption{Plots of $\Omega_\phi$ for Schwarzschild (black solid line), and numerical (dashed lines) and perturbative (dotted lines) MPD-S solutions with the percent deviations (dev) of the perturbative solutions w.r.t. the numerical ones. Left and right panels display $n=2$ and $n=3$ cases, respectively. Color codes and parameter choices are explained in the legends.}
\label{fig:pert}
\end{figure*}

\paragraph*{{\bf \emph{Matching the spin from the disk symmetry.}}} It is now necessary to focus on $\mathcal S^{rt}$ that, from symmetry demands, can be parameterized by a power-law $\propto r^n$, naively corresponding to a filament-like ($n=1$), disk-like ($n=2$), or spherical-like ($n=3$) configurations, as
\begin{equation}
\label{eq12}
\mathcal S^{tr} = \mathcal C_n r^n\,,
\end{equation}
with $\mathcal C_n$ a small amplitude, keeping the model as minimal as possible to guarantee the TPA, i.e.,
\begin{equation}
\label{eq13}
\kappa = |\mathcal C_n|\,|\Omega_\phi|^{-1}
\sqrt{r^{2n-5}(r-2M-r^3\Omega_\phi^2)}\ll1\,.
\end{equation}
Thus, plugging Eq.~\eqref{eqdeltas} into Eq.~\eqref{eq_for_Of} one obtains, 
\begin{align}
\nonumber
&s\,\kappa\,r|\Omega_\phi|\Bigl\{(2M-r)r\partial_r\Omega_\phi
+\bigl[(n-1)r-3M(n-2)\bigr]\Omega_\phi\Bigr\}\\
\label{eq_above}
&\quad+2\Omega_\phi\bigl(r^3\Omega_\phi^2-M\bigr)=0\,,
\end{align}
where $s$ is the sign of $\mathcal C_n$. 
In the innermost regions of the accretion disk, the condition in Eq.~\eqref{eq13} reduces to
\begin{equation}
\label{eq13bis}
\kappa\approx\kappa_0 = |\mathcal C_n|\,r^{n-1}\sqrt{M^{-1}(r-3M)}\ll1\,.
\end{equation}
Accordingly, we solve Eq.~\eqref{eq_above} perturbatively, writing $\Omega_\phi=\Omega_0+\kappa_0\Omega_1+\mathcal O(\kappa_0^2)$, yielding 
\begin{equation}
\label{eq10bis}
\Omega_\phi = \pm\frac{\sqrt{M}}{r^{3/2}}
\mp s\kappa_0\frac{(2n\!+\!1)r-6M(n\!-\!1)}{8r^2}
+ \mathcal O(\kappa_0^2)\,,
\end{equation}
for co- ($+$) and counter-rotating ($-$) orbits,
where the first order in $\kappa_0$ perfectly matches our underlying gauge, in Eq. \eqref{eq2}, moreover agreeing with the numerical solution, within a few per mille, as sketched in Fig.~\ref{fig:pert}. Finally, the radial epicyclic frequency becomes
\begin{equation}
\label{eq11bis}
\Omega_r^2 = \frac{M}{r^3}\left(1-\frac{6M}{r}\right)(1+\delta_{\rm S})\,,
\end{equation}
where \emph{$\delta_{\rm S}$ encodes the spin-induced correction to the Schwarzschild result},
\begin{align}
\nonumber
\delta_{\rm S} =& -\left[M(r-6M)\right]^{-1}\Bigl\{(r-2M)\bigl[3(M-r^3\Omega_\phi^2) \\
\nonumber
&+\Omega_\phi\sqrt{r^3(r-2M-r^3\Omega_\phi^2)}\,
(r\partial_r^2\Delta\mathcal L-4\partial_r\Delta\mathcal L) \\
\nonumber
&-(r-2M-r^3\Omega_\phi^2)(\partial_r\Delta\mathcal L)^2\bigr]\\
\nonumber
&+\sqrt{r^3(r-2M-r^3\Omega_\phi^2)}\,
\bigl[(r-2M)r\partial_r^2\Delta\mathcal E-4M\partial_r\Delta\mathcal E\bigr]\\
\label{eqdelta}
&+(r-2M-r^3\Omega_\phi^2)r^3(\partial_r\Delta\mathcal E)^2\Bigr\}\,.
\end{align}
In the spinless limit, $\Delta\mathcal E=\Delta\mathcal L=0$, the standard RPM is recovered with $\Omega_\phi=\pm\sqrt{M/r^3}$ and $\delta_{\rm S}=0$, see Eq.~\eqref{Ophi_Or_RPM}.

Using Eqs.~\eqref{eqdeltas}--\eqref{eq12}, at first order in $\kappa_0$, one obtains the compact expression
\begin{align}
\nonumber
\delta_{\rm S} = -s\kappa_0\Biggl[&
\frac{3M^2(2n^2-4n-1)-M(5n^2-10n+1)r}{2\sqrt{Mr}(r-6M)}\\
\label{eqdeltaSbis}
&+ \frac{(4n^2-4n-3)r^2}{8\sqrt{Mr}(r-6M)}\Biggr]
+ \mathcal O(\kappa_0^2)\,.
\end{align}

\paragraph*{{\bf \emph{The TPA condition.}}}The TPA shall hold across the disk annulus where QPOs are produced. We denote by $h=\{M,\mathcal C_n\}$ the model parameters and by $f_{{\rm U},k}\pm\sigma_{{\rm U},k}$ the measured upper frequencies for each source. For each $k$ we determine a radius $r_k$ from $f_{{\rm U},k}=f_U(h,r_k)$, and define the inner and outer radii of the above annulus as
\begin{equation}
{\rm max}(f_{{\rm U},k})\equiv f_U(h,r_{\rm in})\ ,\  
{\rm min}(f_{{\rm U},k}) \equiv f_U(h,r_{\rm out})\,,
\end{equation}
leading to radii satisfying
\begin{equation}
\label{eqROC}
r_{\rm ISCO}\leq r_{\rm in}\leq r_{\rm out}\,,
\end{equation}
where $r_{\rm ISCO}$ is the innermost stable circular orbit (ISCO) radius, determined by solving $\Omega_r^2(r_{\rm ISCO})=0$ along with Eq.~\eqref{eq11bis}. From an observational viewpoint, $r_{\rm in}$ ($r_{\rm out}$) turns out to be an upper (lower) limit on the inner (outer) disk edge. The true outer radius is hard to identify observationally, but the TPA offers an additional recipe, i.e., we assume the disk maintains its unaltered properties up to the radius where $\kappa_0\simeq 0.2$. This value corresponds to the absolute upper limit on $\kappa_0$ extracted from the results of the MCMC analyses (see values in Table~\ref{tab:results2}). Accordingly, solving Eq.~\eqref{eq13}, with the latter condition,  defines a \emph{characteristic disk radius}, $r_{\rm disk}$, imposing a corresponding radial ordering condition,
\begin{equation}
\label{eqROC2}
r_{\rm ISCO}\leq r_{\rm in}\leq r_{\rm out}\leq r_{\rm disk}\,.
\end{equation}
This requirement is met for all sources and provides a physically transparent constraint on the inferred disks.


\section{Statistical analysis and results}
\label{sec:statistical}

We now confront our MPM on a Schwarzschild background, hereafter MPD-S, with the QPO data of the eight NS-LMXBs and, for each source, we determine the parameters $h$ and the preferred macroscopic index $n$ by maximizing a log-likelihood function. We thus compare our findings with the SdS and genuine Schwarzschild background through statistical and physical selection criteria, finding evident preference for our paradigm. 

\paragraph*{{\bf \emph{Numerical analysis.}}} Given a set of $N$ twin kHz QPO measurements $(f_{{\rm L},k}\pm\sigma_{{\rm L},k},f_{{\rm U},k}\pm\sigma_{{\rm U},k})$ for a given source, we define the log-likelihood
\begin{equation}
\label{eqloglike}
\ln L = -\frac{1}{2}\sum_{k=1}^{N}
\!\left\{\!
\frac{[f_{{\rm L},k}-f_L(h,r_k)]^2}{\sigma_{{\rm L},k}^2}
+ \ln\!\left(2\pi\sigma_{{\rm L},k}^2\right)\!
\right\},
\end{equation}
where, because of the identity $f_{{\rm U},k}=f_U(h,r_k)$, each radius $r_k$ depends implicitly on $h$. The likelihood thus depends only on the parameters $h$ for a fixed choice of $n$.

We explore the parameter space with the Metropolis-Hastings algorithm, implementing MCMC chains through a \texttt{Wolfram Mathematica} code, with $\mathcal O(10^5)$ iterations per source, by adopting broad priors,
\begin{equation}
M\in[0,5]\,M_\odot\,,\qquad
\mathcal C_n\in[-0.1,0.1]\,{\rm km}^{1-n}\,,
\end{equation}
and considering discrete values of $n$, obtaining for each source the maximum log-likelihood $\ln L_{\rm max}$, and the corresponding parameter posterior distributions.

To compare the performance of different models,  here summarized by the set (S, SdS, MPD-S), we use the Deviance Information Criterion (DIC) \cite{Kunz:2006mc},
\begin{equation}
{\rm DIC} = 2\langle -2\ln L\rangle + 2\ln L_{\rm max}\,,
\end{equation}
where $\langle\cdot\rangle$ denotes an average over the posteriors. The model with the smallest DIC for a given source is taken as reference, and differences $\Delta{\rm DIC}={\rm DIC}-{\rm DIC}_0$ are interpreted as: weak evidence for $0\leq\Delta{\rm DIC}\leq 3$, mild for $3<\Delta{\rm DIC}\leq 6$, and strong for $\Delta{\rm DIC}>6$.

Table~\ref{tab:results} sums up best-fit parameters and statistical analyses. For each NS-LMXB, we list the mass, the SdS curvature parameter $R_0$, the MPD-S spin parameter $\mathcal C_n$ with its sign $s$ and index $n$, and the corresponding $\ln L_{\rm max}$, DIC, and $\Delta{\rm DIC}$. The Schwarzschild and SdS entries are reproduced from Ref.~\cite{boshkayev23b}, obtained there with the same pipeline but without spin corrections.

Two robust features emerge: 
\begin{itemize}
    \item[-] the pure Schwarzschild RPM is strongly disfavored in all sources ($\Delta{\rm DIC}\gg 6$), confirming the tension already highlighted in Ref.~\cite{boshkayev23b}, and
    \item[-] the macroscopic MPD-S model is always competitive with the SdS fits, being
    statistically equivalent ($|\Delta{\rm DIC}|\leq 3$) with it for half catalog of LMXBs, while clearly outperforming it in the other half.
\end{itemize} 

\begin{table*}
\footnotesize
\centering
\setlength{\tabcolsep}{.7em}
\renewcommand{\arraystretch}{1.3}
\begin{tabular}{llrrrrrrrr}
\hline\hline
Source &                      Spacetime &  
\multicolumn{5}{c}{Best-fit parameters} &
\multicolumn{3}{c}{Statistics}          \\
&                                       &
$M\ ({\rm M}_\odot)$                    & 
$R_0/10^{-5}\,({\rm km}^{-2})$          &
$n$                                     &
$s$                                     &
$\log(\mathcal C_n/{\rm km}^{-n+1})$    &
$\ln L_{\rm max}$                       &
DIC                                     &
$\Delta{\rm DIC}$                       \\
\hline\hline
Cir X1                                  &
S                                       &
$2.224^{+0.029\,(0.058)}_{-0.029\,(0.058)}$ & 
--                                      &
--                                      &
--                                      &
--                                      &
$-125.84$ & $254$ & $118$ \\
                                        &
SdS                                     &
$1.846^{+0.045\,(0.091)}_{-0.045\,(0.090)}$ & 
$1.28^{+0.12\,(0.23)}_{-0.12\,(0.24)}$  &
--                                      &
--                                      &
--                                      &
$-70.07$ & $144$ & $8$\\
                                        &
MPD-S                                   &
$1.283^{+0.056\,(0.097)}_{-0.058\,(0.097)}$  &
--                                      &
$2$                                     &
$+1$                                     &
$-2.742^{+0.022\,(0.036)}_{-0.030\,(0.048)}$ &
$-65.96$ & $136$ & $0$ \\
\hline
GX 5--1                                 &
S                                       &
$2.161^{+0.010\,(0.020)}_{-0.010\,(0.021)}$ & 
--                                      &
--                                      &
--                                      &
--                                      &
$-200.33$ & $403$ & $187$ \\
                                        &
SdS                                     &
$2.397^{+0.019\,(0.038)}_{-0.019\,(0.038)}$ & 
$-6.46^{+0.48\,(0.95)}_{-0.48\,(0.95)}$ &
--                                      &
--                                      &
--                                      &
$-106.08$ & $217$ & $1$\\
                                        &
MPD-S                                   & 
$2.427^{+0.030\,(0.050)}_{-0.031\,(0.051)}$ &
--                                      &
$3$                                     &
$-1$                                     &
$-5.042^{+0.047\,(0.080)}_{-0.045\,(0.069)}$ &
$-105.73$ & $216$ & $0$\\
\hline
GX 17+2                                 & 
S                                       &
$2.077^{+0.001\,(0.002)}_{-0.001\,(0.002)}$  &
--                                      &
--                                      &
--                                      &
--                                      &
$-1819.02$ & $3642$ & $3544$ \\
                                        &
SdS                                     &
$1.733^{+0.011\,(0.021)}_{-0.011\,(0.022)}$  &
$21.53^{+0.45\,(0.91)}_{-0.45\,(0.90)}$ & 
--                                      &
--                                      &
--                                      &
$-46.42$ & $98$ & $0$ \\
                                        &
MPD-S                                   &
$1.691^{+0.015\,(0.025)}_{-0.017\,(0.027)}$  &
--                                      &
$3$                                     &
$+1$                                     &
$-4.514^{+0.013\,(0.021)}_{-0.015\,(0.024)}$ &
$-46.56$ & $98$ & $0$ \\
\hline
GX 340+0                                &
S                                       &
$2.102^{+0.003\,(0.007)}_{-0.003\,(0.007)}$  & 
--                                      &
--                                      &
--                                      &
--                                      &    
$-130.86$ & $264$ & $10$ \\
                                        &
SdS                                     &
$2.149^{+0.015\,(0.030)}_{-0.015\,(0.031)}$  & 
$-1.39^{+0.45\,(0.89)}_{-0.44\,(0.89)}$ &
--                                      &
--                                      &
--                                      &
$-126.06$ & $257$ & $3$ \\
                                        &
MPD-S                                   &
$2.274^{+0.066\,(0.120)}_{-0.049\,(0.099)}$  &
--                                      &
$2$                                     &
$-1$                                     &
$-3.312^{+0.142\,(0.366)}_{-0.139\,(0.225)}$ & 
$-124.78$ & $254$ & $0$\\
\hline
Sco X1                                  &
S                                       &
$1.965^{+0.001\,(0.002)}_{-0.001\,(0.002)}$ & 
--                                      &
--                                      &
--                                      &
--                                      &
$-3887.17$ & $7776$ & $7499$\\
                                        &
SdS                                     &    $1.690^{+0.003\,(0.007)}_{-0.003\,(0.007)}$ & 
$21.77^{+0.24\,(0.49)}_{-0.25\,(0.49)}$ &
--                                      &
--                                      &
--                                      &
$-158.61$ & $323$ & $46$ \\                 
                                        &
MPD-S                                   &
$1.372^{+0.006\,(0.013)}_{-0.007\,(0.013)}$ &
--                                      &
$2$                                     &
$+1$                                     &
$-2.527^{+0.005\,(0.010)}_{-0.005\,(0.011)}$ &
$-136.72$ & $277$ & $0$ \\
\hline
4U1608--52 & S                          &
$1.960^{+0.004\,(0.007)}_{-0.004\,(0.008)}$ & 
--                                      & 
--                                      &
--                                      &
--                                      &
$-235.83$ & $474$ & $345$\\
        & SdS                           &
$1.728^{+0.014\,(0.028)}_{-0.014\,(0.028)}$ &
$17.62^{+0.94\,(1.87)}_{-0.94\,(1.88)}$ &
--                                      &
--                                      &
--                                      &
$-66.14$ & $137$ & $8$ \\
        & MPD-S                         & 
$1.429^{+0.027\,(0.052)}_{-0.037\,(0.060)}$ &
--                                      &
$2$                                     &
$+1$                                     &
$-2.584^{+0.030\,(0.049)}_{-0.025\,(0.048)}$ &
$-62.32$ & $129$ & $0$ \\
\hline
4U1728--34 & S                          &
$1.734^{+0.003\,(0.006)}_{-0.003\,(0.006)}$ & 
--                                      & 
--                                      & 
--                                      & 
--                                      &
$-212.61$ & $427$ & $353$ \\
           & SdS                        &
$1.445^{+0.016\,(0.032)}_{-0.016\,(0.032)}$ & 
$30.74^{+1.58\,(3.15)}_{-1.58\,(3.18)}$ &
--                                      & 
--                                      & 
--                                      &
$-35.15$ & $74$ & $0$ \\
                                        & 
MPD-S                                   &
$1.115^{+0.032\,(0.059)}_{-0.030\,(0.056)}$ & 
--                                      & 
$2$                                     &
$+1$                                     &
$-2.404^{+0.024\,(0.043)}_{-0.025\,(0.048)}$ &
$-35.02$ & $74$ & $0$ \\
\hline
4U0614+091                              & 
S                                       &
$1.904^{+0.001\,(0.003)}_{-0.001\,(0.003)}$ & 
--                                      & 
--                                      &
--                                      & 
--                                      &
$-842.97$ & $1670$ & $1345$ \\
                                        & 
SdS                                     &
$1.545^{+0.011\,(0.021)}_{-0.011\,(0.021)}$ & 
$28.39^{+0.80\,(1.59)}_{-0.80\,(1.59)}$ &
--                                      &
--                                      &
--                                      &
$-188.70$ & $382$ & $37$ \\
                                        &
MPD-S                                   &
$1.154^{+0.019\,(0.036)}_{-0.020\,(0.037)}$ &
--                                      & 
$2$                                     &
$+1$                                     &
$-2.397^{+0.013\,(0.024)}_{-0.013\,(0.025)}$ &  
$-170.17$ & $345$ & $0$  \\
\hline\hline
\end{tabular}
\caption{Results of the MCMC fits. The first two columns list the source and the spacetime. The next five columns show the best-fit parameters with $1\sigma$ ($2\sigma$) errors. The last three columns report the maximum log-likelihood, the DIC, and the difference $\Delta{\rm DIC}$ w.r.t. the best model for each source. Values for S and SdS (in a couple of cases Schrwarzschild -- anti-de Sitter) are taken from Ref.~\cite{boshkayev23b}, obtained with the same pipeline used here for MPD-S.}
\label{tab:results} 
\end{table*}

Remarkably, in all NS-LMXBs the data single out the macroscopic indices in the narrow range $2\leq n\leq3$, whereas no statistically acceptable solution is found for $n=1$, placing the majority to prefer $n=2$. For the sake of clearness, we restrict our analysis to discrete $n$ values, since these seem to appear  statistically preferred over continuous values $n\in[1,3]$.

Remarkably, the “near-degeneracy” between SdS and MPM frameworks predicts a cosmological constant-like behavior that {\bf 1)} mimics a genuine cosmological contribution entirely arising from spin effects, {\bf 2)} does not contradict the \emph{Kerr hypothesis}, since compact objects may still be consistently described by either the Schwarzschild or the Kerr solutions~\cite{Mazur:2004fk}, and {\bf 3)} remains applicable to \emph{any spacetimes}, explicitly leaving open the chance to adapt to mimickers~\cite{Bambi:2025wjx} or other exotic configurations~\cite{Luongo:2025iqq}.

This can be clearly understood by comparing the leading corrections to the radial epicyclic frequency at large radii. Both $R_0$ and $\mathcal C_n$ must be clearly very small to guarantee the TPA to hold, as certified by the fits. Accordingly, for small $R_0$ and  $\mathcal C_n$, one immediately finds
\begin{equation}
\label{SdS_vs_MPDS}
\begin{aligned}
\delta_{\rm SdS} &\approx -\frac{R_0 r^3}{3M},\\
\delta_{\rm S} &\approx -\frac{\mathcal C_n(4n^2 - 4n - 3)\, r^n}{8M}\ 
\xRightarrow[\;n=3\;]{}\ 
\mathcal C_n \simeq \frac{8R_0}{63}\,,
\end{aligned}
\end{equation}
providing a rough estimate of $C_n$ and implying that it effectively mimics a cosmological constant. However, Table~\ref{tab:results} exhibits preference for $n=2$ and thus if  $\delta_{\rm SdS}-\delta_{\rm S}\simeq 0$, then from Eq.~\eqref{SdS_vs_MPDS} it follows that $\mathcal C_n \simeq 8 r_{\rm disk} R_0/15$. So, imposing the extreme and unrealistic case $|R_0|\simeq r_{\rm disk}^{-2}$, one immediately finds $|\mathcal C_n|\sim 1/(2r_{\rm disk})$, suggesting that for realistic $r_{\rm disk}$, $\delta_{\rm SdS}$ and $\delta_{\rm S}$ appear quite similar.

Figure~\ref{fig:freq} shows, for all eight sources, the observed frequency pairs and the best-fitting curves for S, SdS, and MPD-S, together with the residuals, computed w.r.t. MPD-S. While the pure Schwarzschild curves systematically depart from the data, both SdS and MPD-S track the observed trends very closely. 
The MPM provides slightly better or comparable fits, and offers a clear physical interpretation in terms of spin-curvature coupling rather than an effective and unmotivated $R_0$ term.

\begin{figure*}
\includegraphics[width=0.47\hsize,clip]{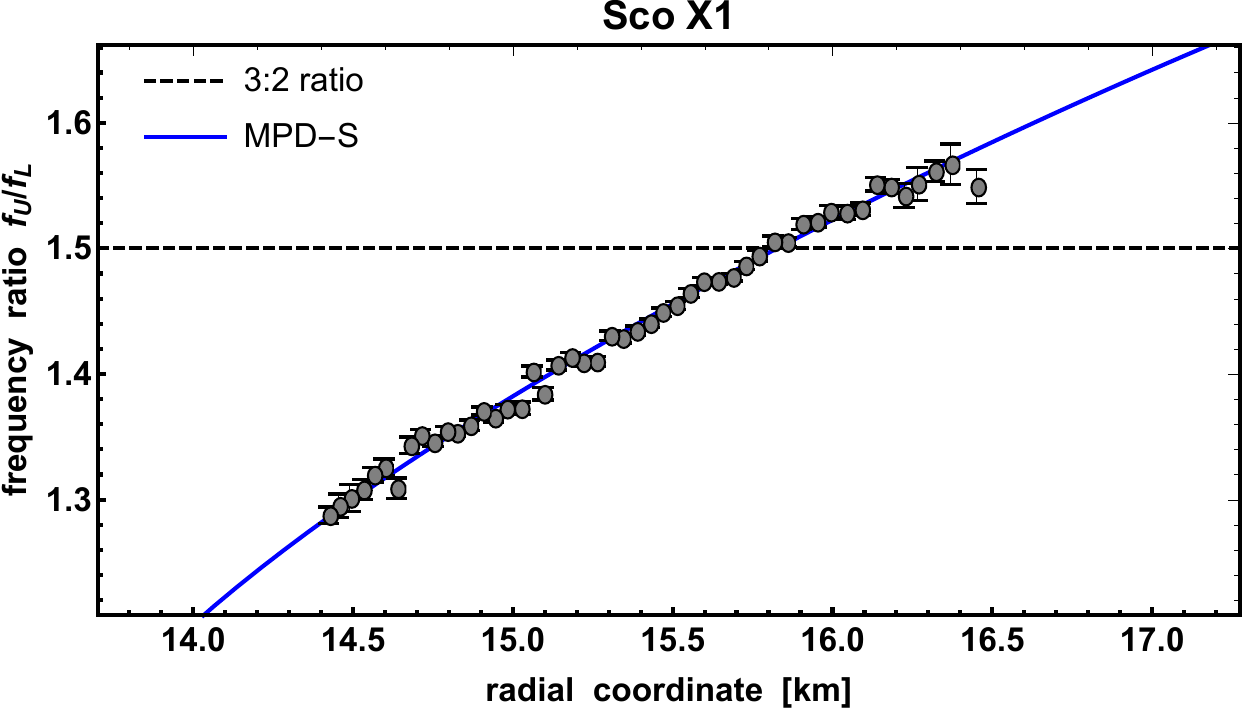}
\hfill
\includegraphics[width=0.47\hsize,clip]{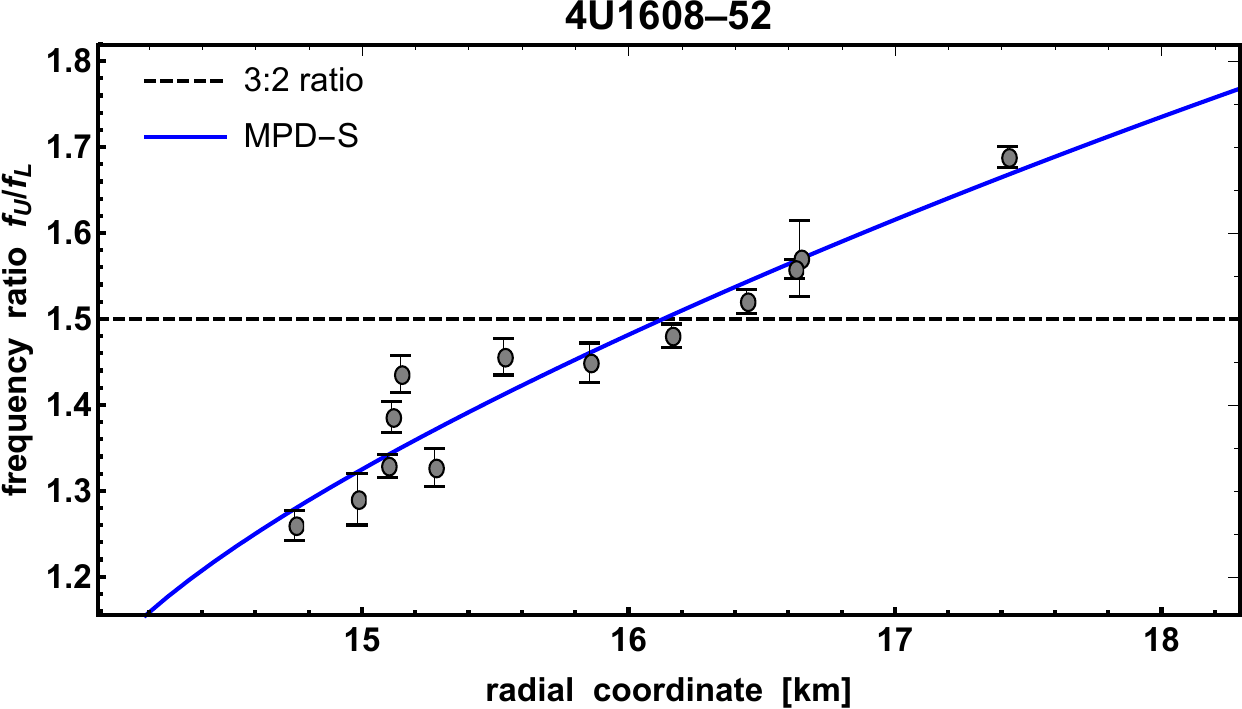}\\
\vspace{0.2cm}
\includegraphics[width=0.47\hsize,clip]{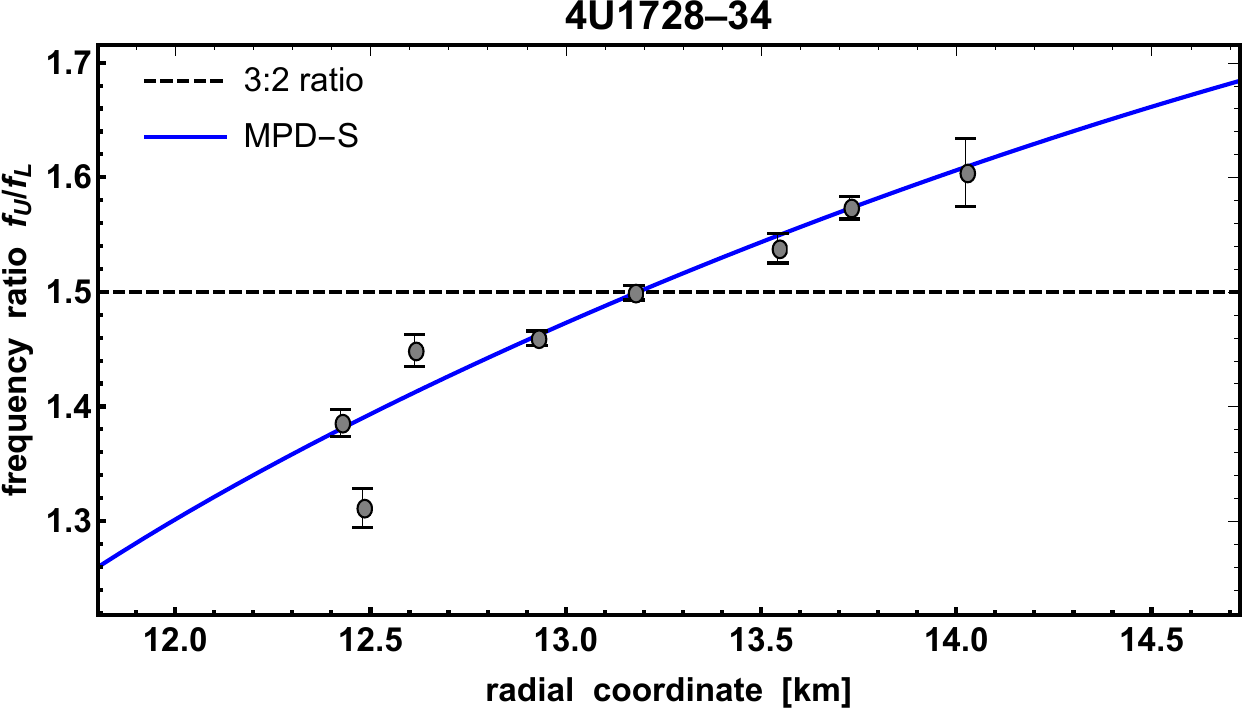}
\hfill
\includegraphics[width=0.47\hsize,clip]{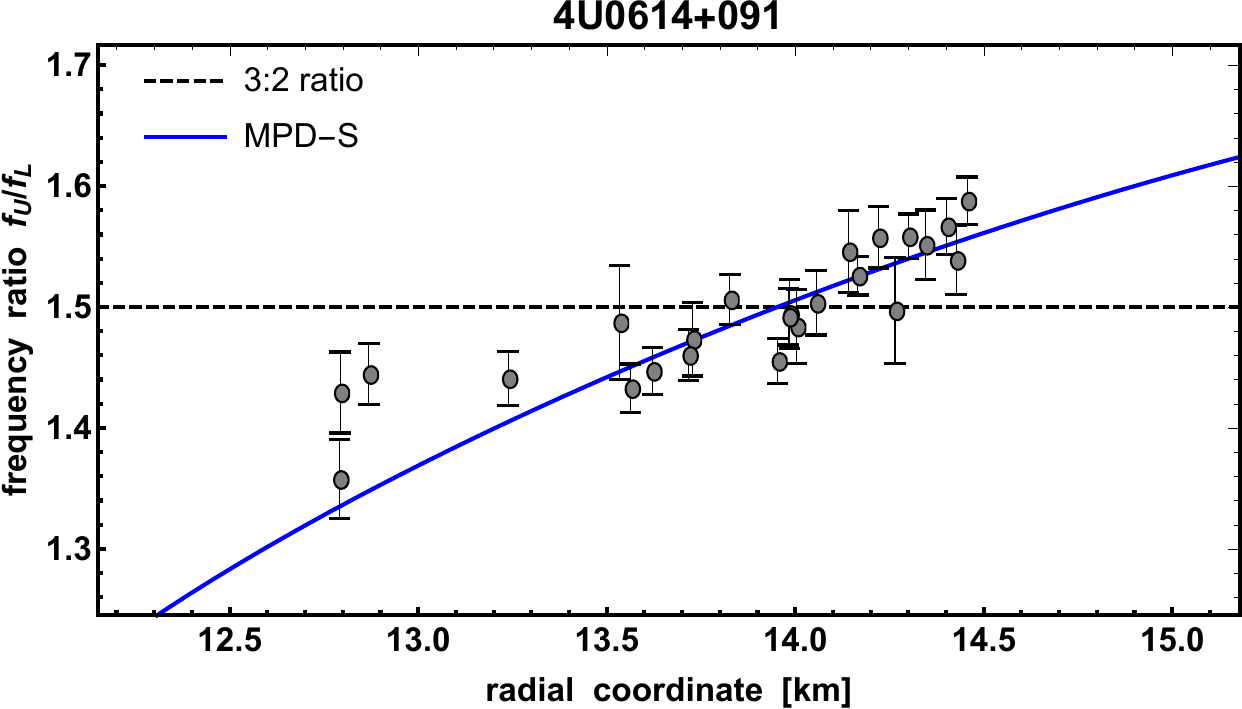}
\caption{Forecasts of the $3:2$ ratio approximately obtained by our MPM in four LMXBs with compact disks (see Table~\ref{tab:results2}).}
\label{fig:ratio}
\end{figure*}

\paragraph*{{\bf \emph{Consistency check on masses and disks.}}} Beyond statistical preference, the MPM model requires:
\begin{itemize}
\item[-] NS masses no larger than the theoretical upper bound $M\lesssim 3.2\,M_\odot$ \cite{2002BASI...30..523S},
\item[-] TPA validity, viz. $\kappa_0\ll1$,  throughout all radii probed by the QPOs, and
\item[-] the ordering of characteristic radii, see Eq.~\eqref{eqROC2}.
\end{itemize}

Table~\ref{tab:results2} reports the inferred $r_{\rm ISCO}$, $r_{\rm in}$, $r_{\rm out}$, and $r_{\rm disk}$, along with the values of $\kappa$ at $r_{\rm in}$ and $r_{\rm out}$ for the MPM case; for Schwarzschild and SdS cases, the results are taken from Ref.~\cite{boshkayev23b}.
For all eight sources the macroscopic model satisfies $r_{\rm ISCO}\leq r_{\rm in}\leq r_{\rm out}\leq r_{\rm disk}$ and yields TPA parameters in the range $(\kappa_{\rm in},\kappa_{\rm out})\sim[0.01$-$0.2]$, well within the test particle regime, moreover showing NS masses comfortably below $3.2\,M_\odot$.


\section{Final outlooks}
\label{sec:conclusions}

We revisited the RPM including the internal structure of test bodies through the MPD formalism and, in so doing, proposed a new paradigm dubbed MPM, whose main results are summarized as follows.

\begin{itemize}
\item[-] The MPM applied to a Schwarzschild background modifies both the azimuthal and the radial epicyclic frequencies, quantifying a trend for test particles to exhibit internal structure.

\item[-] The spin $\mathcal S^{tr}=\mathcal C_n r^n$ exhibits statistical preference for a disk-like symmetry ($n=2$) and, just in two cases, for a spherical one ($n=3$). We found no evidences for Taylor and $n=1$ reconstructions, or inverse power law terms in the metric functions, namely additional quadrupoles.

\item[-] Frequency corrections demonstrated that the role of $R_0$ is only phenomenological. Accordingly, the MPM is overall statistically preferred than SdS, or at most indistinguishable in some cases. 

\item[-] The sign of $\mathcal C_n$ allows flipping between co- and counter-rotating macroscopic modes, like in de Sitter/anti-de Sitter cases, albeit in this case with a precise physical interpretation associated with spin.

\item[-] The TPA parameter provided a natural way to delineate the radial extent of the disk, exhibiting $\kappa\lesssim 0.2$ at the outer edge, suggesting a simple physical scheme where QPOs may be affected by spin-curvature effects. 
As consequence, no \emph{ad hoc} deformations of the gravitational sector are needed.

\item[-] The inferred masses lied on viable ranges, appearing quite compatible with previous literature \cite{Boutloukos:2006ts,Cherepashchuk:2021ftr,Guver:2008gc,Shaposhnikov:2003cw,vanDoesburgh:2018oom} and, then, showed that the MPM can be effectively used to predict bounds on compact objects. 

\item[-] Finally, the $3:2$ frequency ratio is approximately obtained and shown in Fig.~\ref{fig:ratio} for four LMXBs with compact disks (see Table~\ref{tab:results2}) as a natural evidence for internal structure in test bodies.

\end{itemize}

In view of all the aforementioned considerations, the MPM offered a viable robust alternative to the RPM, moreover being a serious candidate to explain the origins of QPOs, as consequence of the spin-coupled with curvature precession of test bodies with internal structure. 

Even though the model appeared highly predictive, some challenges may remain, i.e., ${\bf a)}$ the spin dependence may be alternatively reconstructed, i.e., more sources are thus essential; ${\bf b)}$ the use of the Kerr metric will represent the next step, to quantify NS quadrupole contributions. Remarkably, even in this case the RPM was not able to be predictive \cite{Boshkayev:2023esw}; ${\bf c)}$ backreaction  and history forces, usually connected to  spin corrections, will need additional investigation to better characterize the disk structure.


\section*{Acknowledgements} 
OL acknowledges support by the Fondazione ICSC, Spoke 3 Astrophysics and Cosmos Observations. National Recovery and Resilience Plan (Piano Nazionale di Ripresa e Resilienza, PNRR) Project ID $CN00000013$ ``Italian Research Center on High-Performance Computing, Big Data and Quantum Computing'' funded by MUR Missione 4 Componente 2 Investimento 1.4: Potenziamento strutture di ricerca e creazione di ``campioni nazionali di R\&S (M4C2-19)'' - Next Generation EU (NGEU). 
OL is particularly grateful for the continuous support offered by Maryam Azizinia during the time in which this work has been realized and expresses his gratitude to Vincenzo Antonuccio-Delogu, Edmund Copeland, Roberto Giambò, Hernando Quevedo and Bharat Ratra for heated debates on related-subjects. OL and MM are also grateful to Stefano Mancini for insightful discussion. The data, employed in this work, have been taken from the recent publications \cite{boshkayev23b,boshkayev23a} and provided by Kuantay Boshkayev and Mariano Mendez who are warmly acknowledged. 
    
\bibliography{bibliography}


\appendix*

\section{Numerical results}

Table~\ref{tab:results2} reports the values of the TPA parameters and the characteristic radii of the accretion disks of the eight LMXBs considered in this work. 
Fig.~\ref{fig:freq} portraits the corresponding best-fit curves with the residuals. 
Fig.~\ref{fig:MCMC} sketches the contours obtained from MCMC analyses, with best fit prompted by the black dots and the associated $2\sigma$ confidence levels.

\setcounter{table}{0}
\renewcommand{\thetable}{A-\arabic{table}}

\begin{table*}
\footnotesize
\centering
\setlength{\tabcolsep}{2.em}
\renewcommand{\arraystretch}{.7}
\begin{tabular}{llcccccc}
\hline\hline
Source                                  &
Spacetime                               & 
$r_{\rm ISCO}$                          &  
$r_{\rm in}$                            & 
$r_{\rm out}$                           &
$r_{\rm disk}$                          &
$\kappa_{\rm in}$                       &
$\kappa_{\rm out}$                      \\
& & (km) & (km) & (km) & (km) & &       \\
\hline\hline
Cir X1  & S                             &
$19.62$ & $30.79$ & $52.16$ & $-$       & 
$-$ & $-$ \\
        & SdS                           &
$16.32$ & $28.84$ & $48.29$ & $-$       & 
$-$ & $-$ \\
        & MPD-S                         & 
$11.92$ & $21.74$ & $29.60$ & $30.44$   & 
$0.12$ & $0.19$ \\
\hline
GX 5-1 & S                              &
$19.06$ & $21.33$ & $31.70$ & $-$       & 
$-$ & $-$ \\
        & SdS                           &
$20.73$ & $22.15$ & $33.35$ & $-$       & 
$-$ & $-$                               \\
        & MPD-S                         &
$20.91$ & $22.32$ & $34.20$ & $72.68$   & 
$0.01$ & $0.03$                         \\
\hline
GX 17+2 & S                             &
$18.32$ & $18.33$ & $22.94$ & $-$       & 
$-$ & $-$ \\
        & SdS                           &
$16.01$ & $17.04$ & $21.09$ & $-$       & 
$-$ & $-$                               \\
        & MPD-S                         &
$15.78$ & $16.79$ & $20.56$ & $41.89$   & 
$0.02$ & $0.03$                         \\
\hline
GX 340+0 & S                            &    
$18.54$ & $21.52$ & $29.07$ & $-$       & 
$-$ & $-$                               \\
        & SdS                           &
$18.88$ & $21.71$ & $29.34$ & $-$       & 
$-$ & $-$                               \\
        & MPD-S                         & 
$19.67$ & $22.51$ & $31.09$ & $86.09$   &
$0.02$ & $0.04$                         \\
\hline
Sco X1  & S                             &
$17.33$ & $17.72$ & $20.98$             & 
$-$ & $-$ & $-$ \\ 
        & SdS                           &
$15.60$ & $16.66$ & $19.58$             & 
$-$ & $-$ & $-$\\
        & MPD-S                         & 
$13.49$ & $14.42$ & $16.45$ & $23.13$   &
$0.09$ & $0.11$ \\
\hline
4U1608--52 & S                         &
$17.29$ & $17.65$ & $21.75$ & $-$      & 
$-$ & $-$ \\ 
        & SdS                          &
$15.82$ & $16.77$ & $20.50$ & $-$      & 
$-$ & $-$ \\
        & MPD-S                      & 
$13.87$ & $14.75$ & $17.42$ & $25.43$  &
$0.08$ & $0.10$ \\
\hline
4U1728--34 & S                         &
$15.30$ & $16.06$ & $18.93$ & $-$      & 
$-$ & $-$ \\ 
        & SdS                          &
$13.37$ & $14.96$ & $17.43$ & $-$      & 
$-$ & $-$ \\
        & MPD-S                        & 
$11.11$ & $12.42$ & $14.02$ & $17.98$  &
$0.10$ & $0.13$ \\
\hline
4U0614+091 & S                         &
$16.80$ & $16.95$ & $20.05$ & $-$      & 
$-$ & $-$ \\ 
        & SdS                          &
$14.34$ & $15.60$ & $18.30$ & $-$      & 
$-$ & $-$ \\
        & MPD-S                      & 
$11.63$ & $12.79$ & $14.45$ & $18.05$  &
$0.11$ & $0.14$ \\
\hline\hline
\end{tabular}
\caption{Derived disk radii for the different models. The columns list, respectively, the source and the spacetime, the ISCO radius, the inner and outer radii deduced from the data points, and the characteristic disk radius $r_{\rm disk}$ where $\kappa_0\simeq0.2$. The last two columns show the TPA parameter at $r_{\rm in}$ and $r_{\rm out}$ for MPD-S. Schwarzschild and SdS values are taken from Ref.~\cite{boshkayev23b}.}
\label{tab:results2} 
\end{table*}

\setcounter{figure}{0}
\renewcommand{\thefigure}{A-\arabic{figure}}

\begin{figure*}
{\hfill
\includegraphics[width=0.48\hsize,clip]{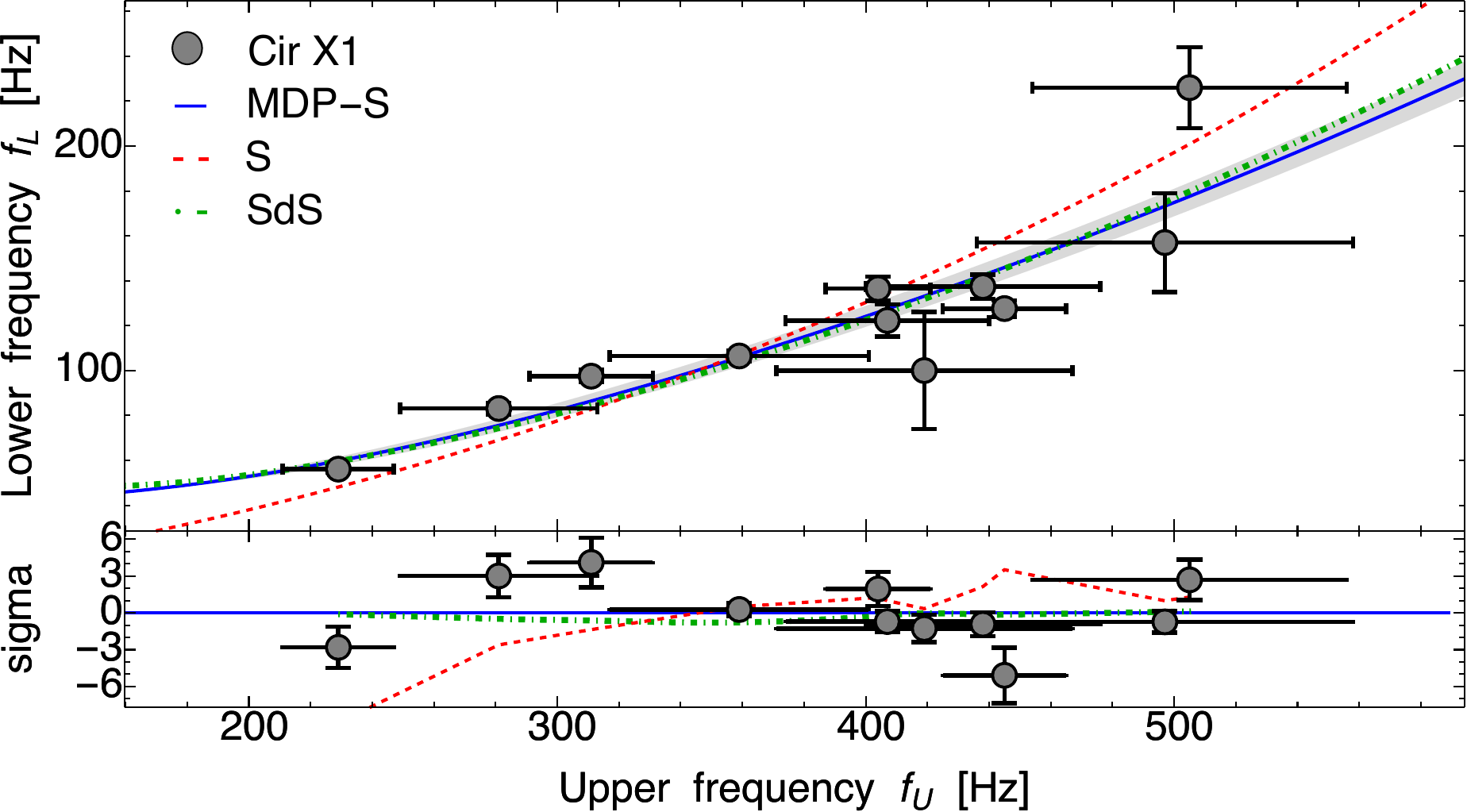}\hfill
\includegraphics[width=0.48\hsize,clip]{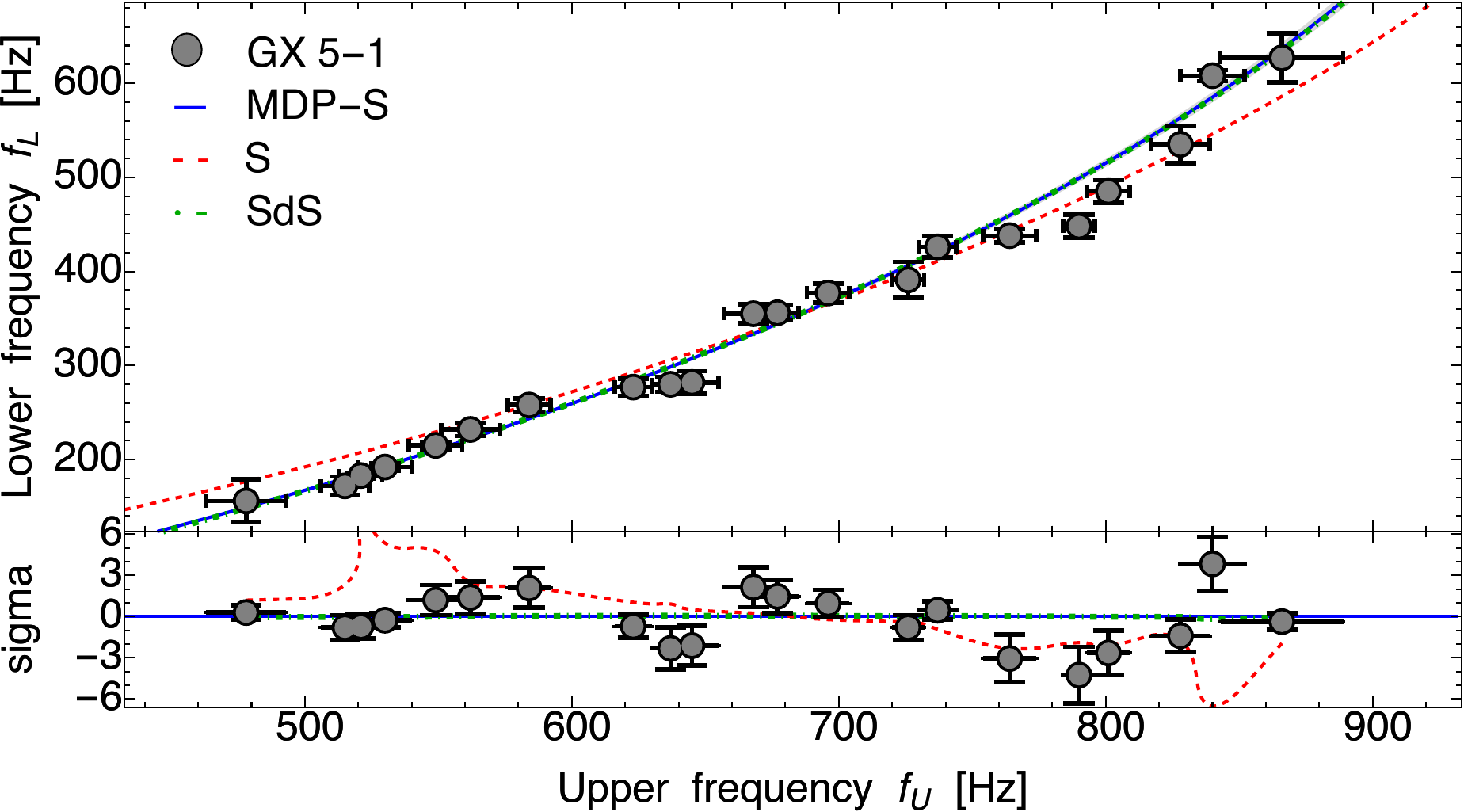}\hfill}\\
\vspace{0.05cm}
{\hfill
\includegraphics[width=0.48\hsize,clip]{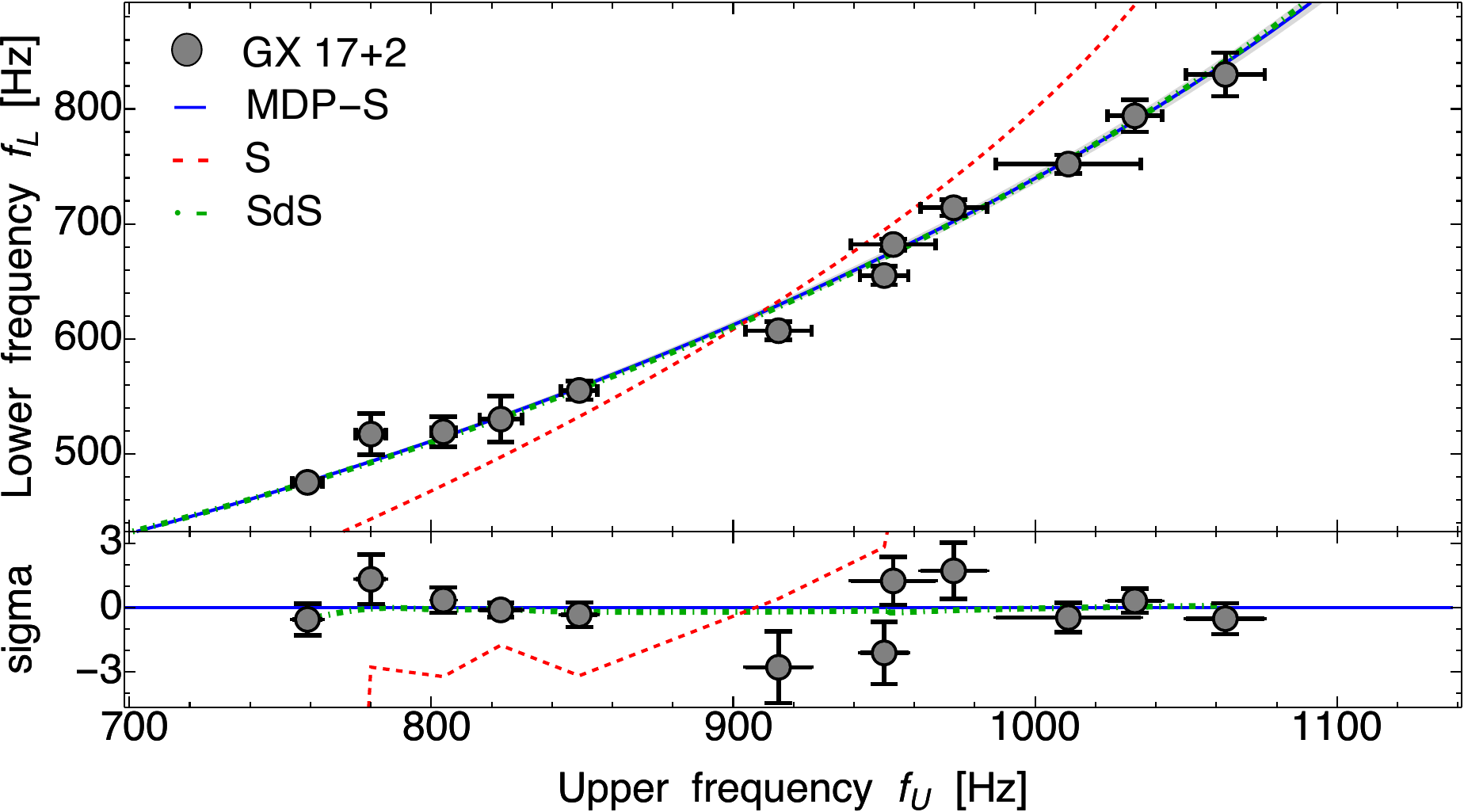}\hfill
\includegraphics[width=0.48\hsize,clip]{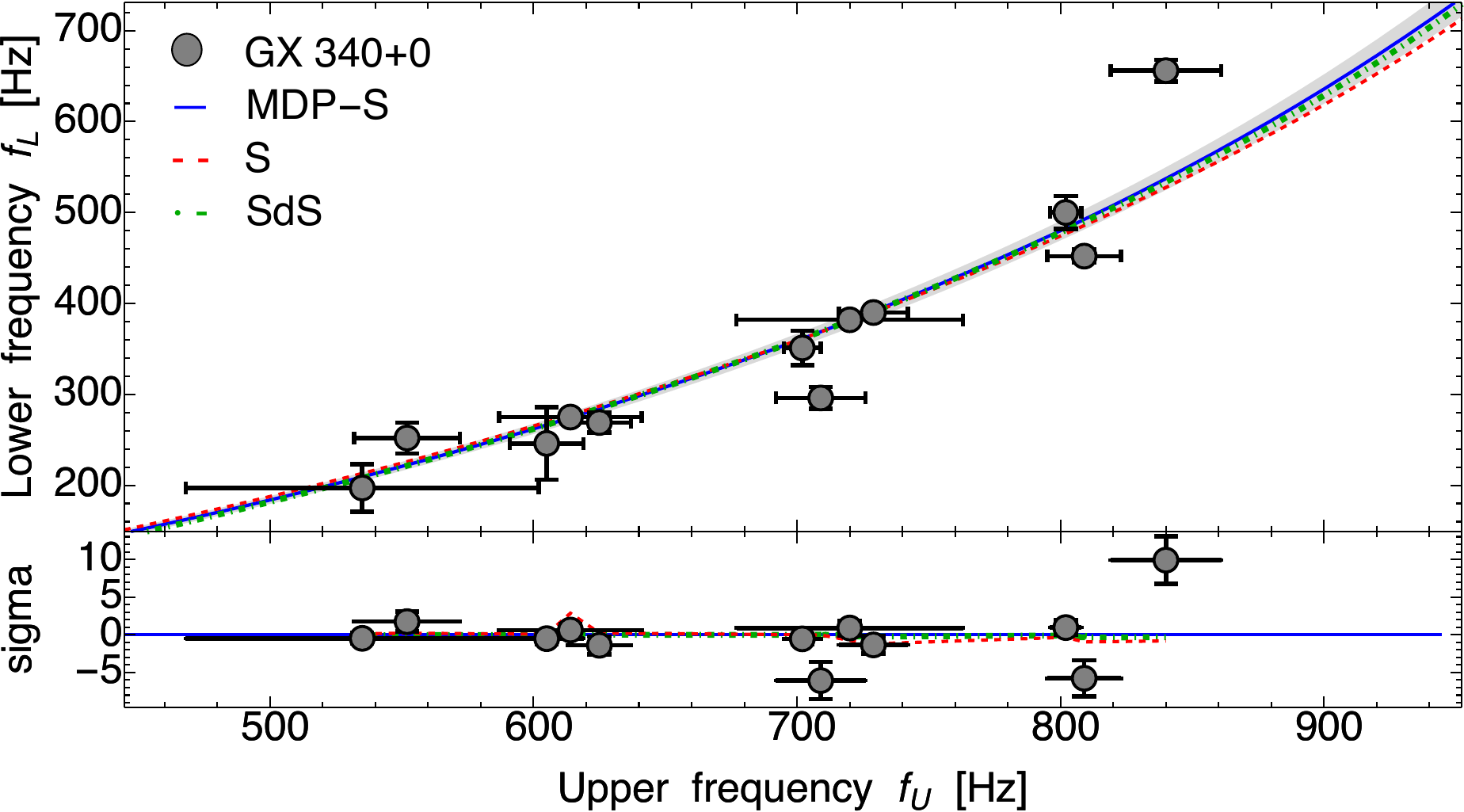}\hfill}\\
\vspace{0.05cm}
{\hfill
\includegraphics[width=0.48\hsize,clip]{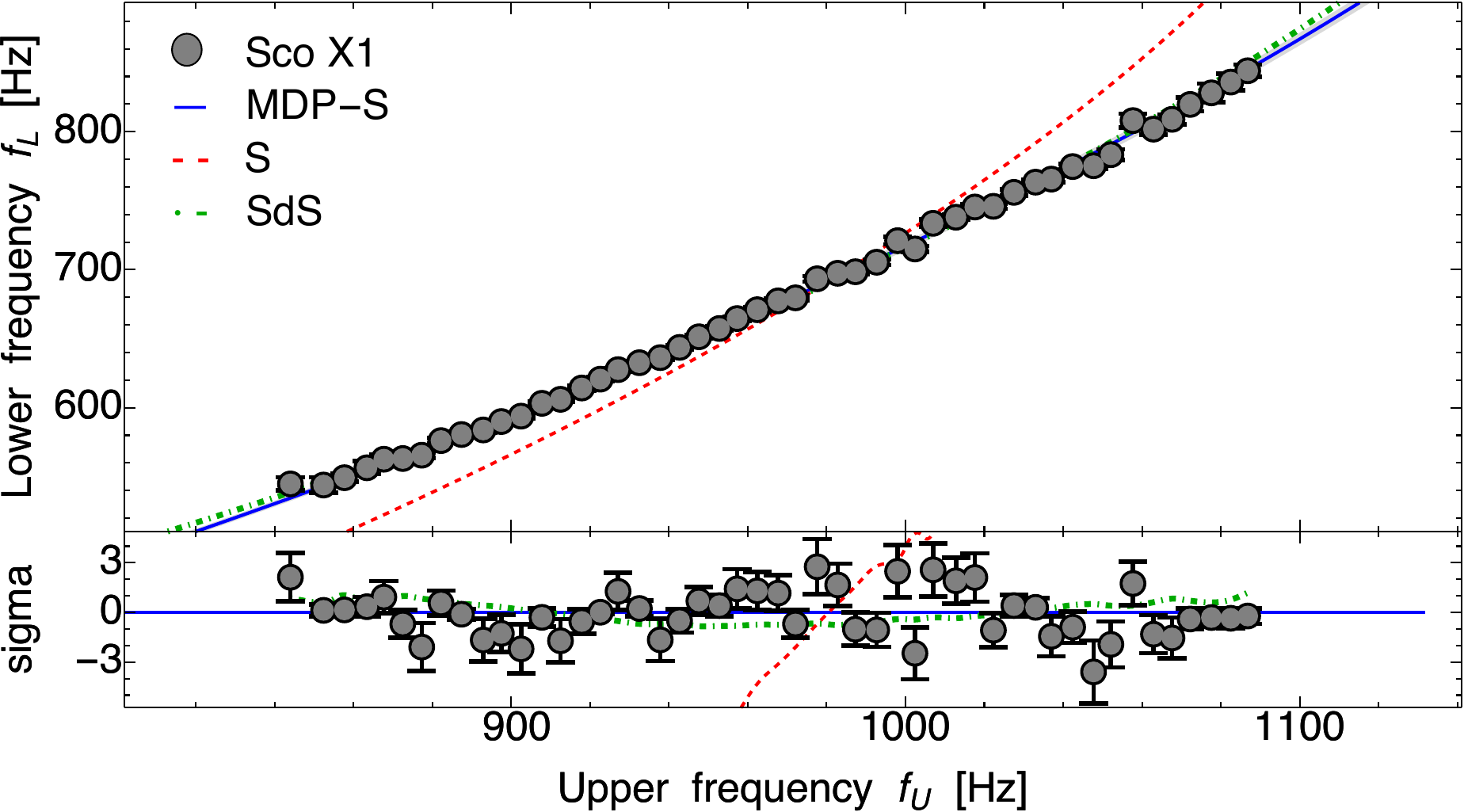}\hfill
\includegraphics[width=0.48\hsize,clip]{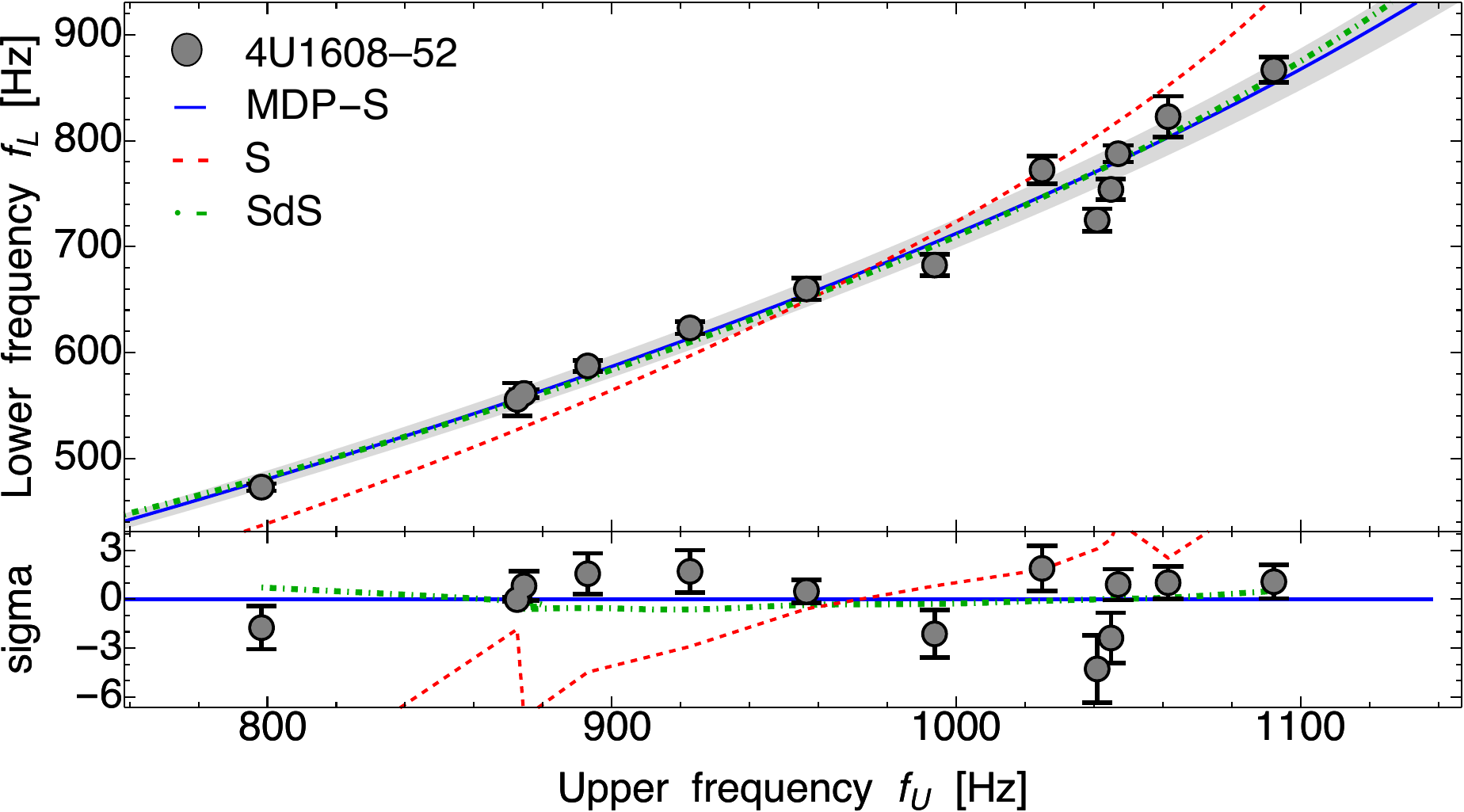}\hfill}\\
\vspace{0.05cm}
{\hfill
\includegraphics[width=0.48\hsize,clip]{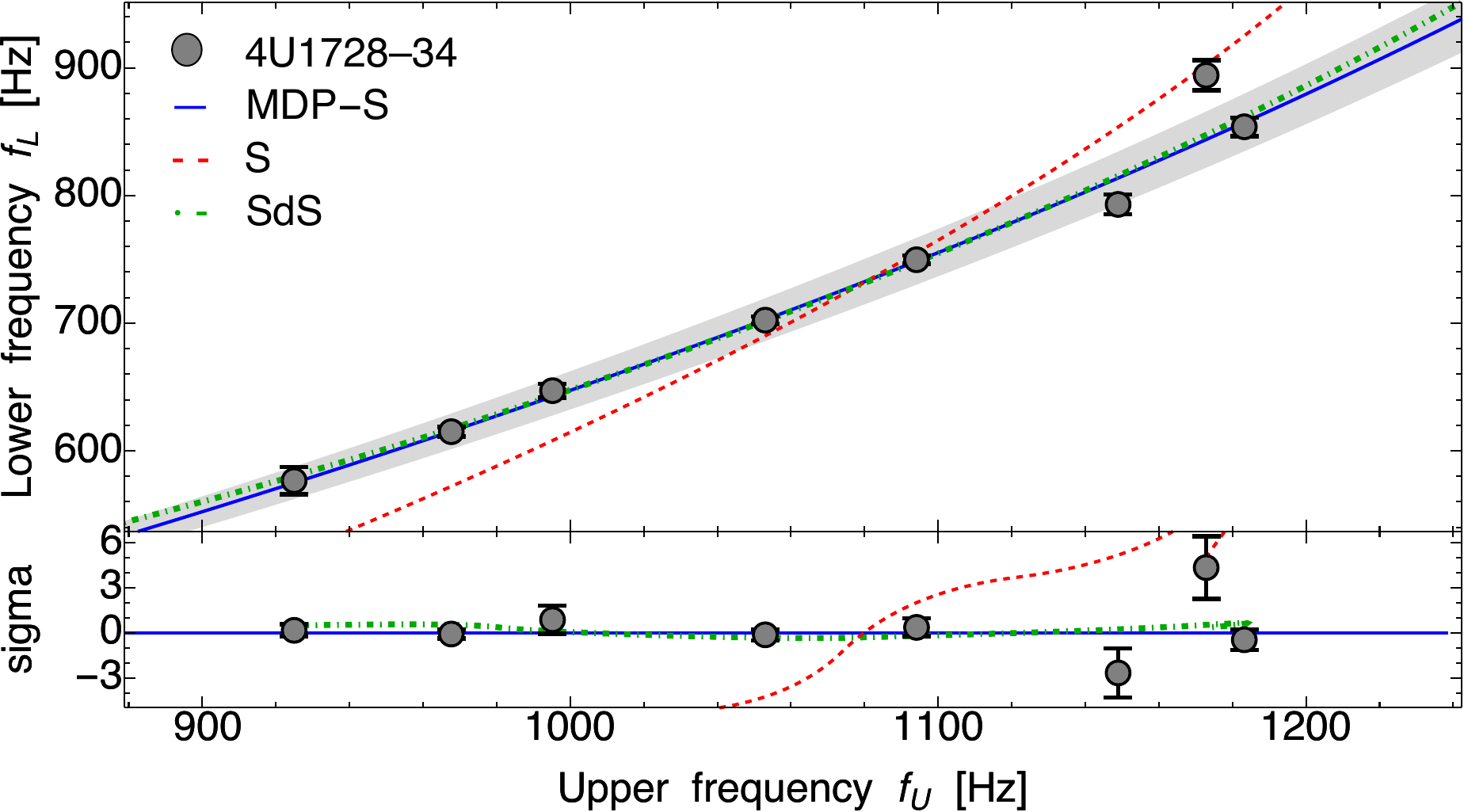}\hfill
\includegraphics[width=0.48\hsize,clip]{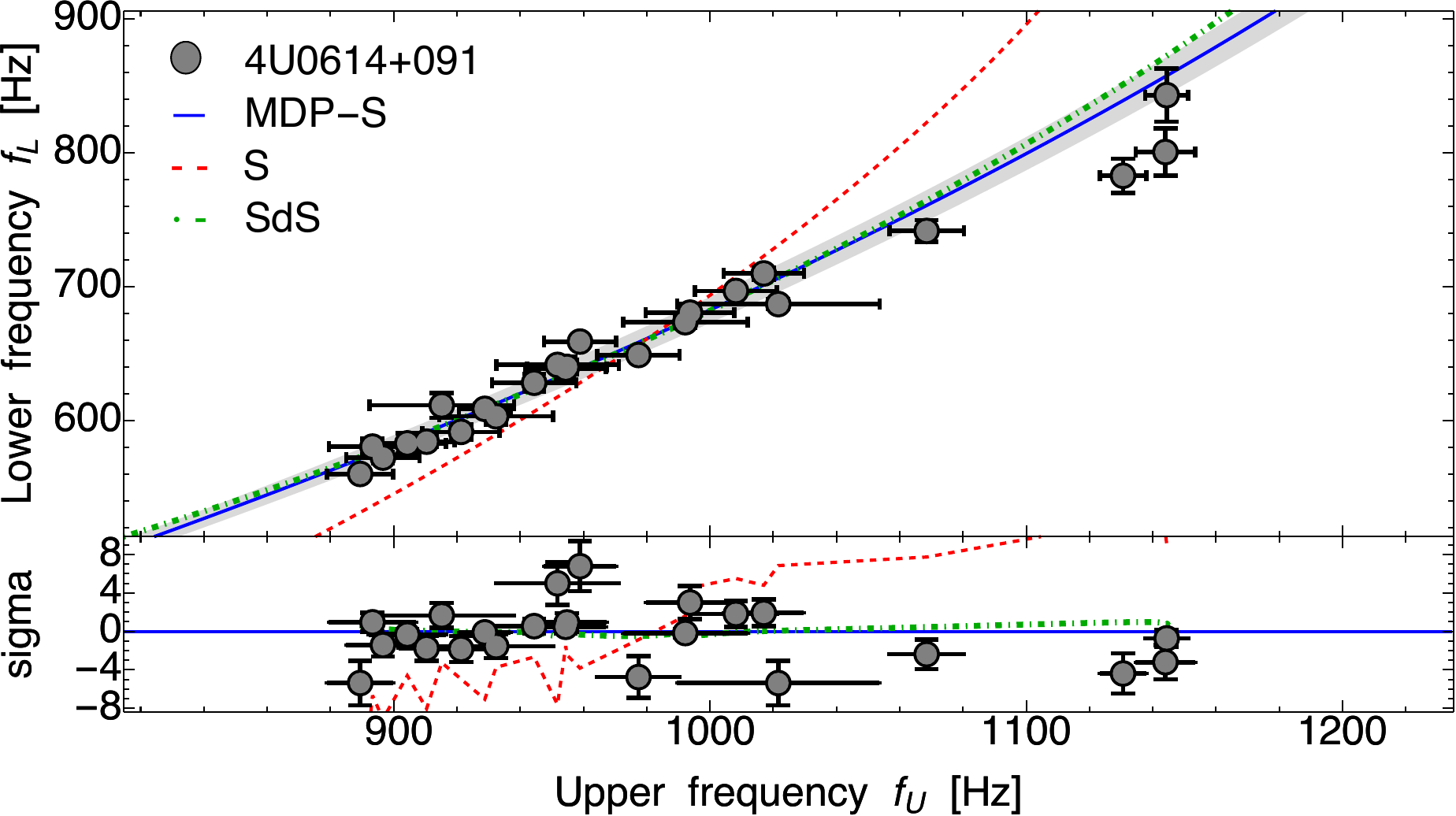}\hfill}\\
\caption{Frequency pairs $(f_{\rm L},f_{\rm U})$ for the eight NS-LMXBs and best-fitting curves for Schwarzschild (dashed red), SdS (dot-dashed green), and MPD-S (solid blue with shaded $1\sigma$ bands). Lower panels show residuals w.r.t. the MPD-S model.}
\label{fig:freq}
\end{figure*}

\begin{figure*}
\includegraphics[width=0.325\hsize,clip]{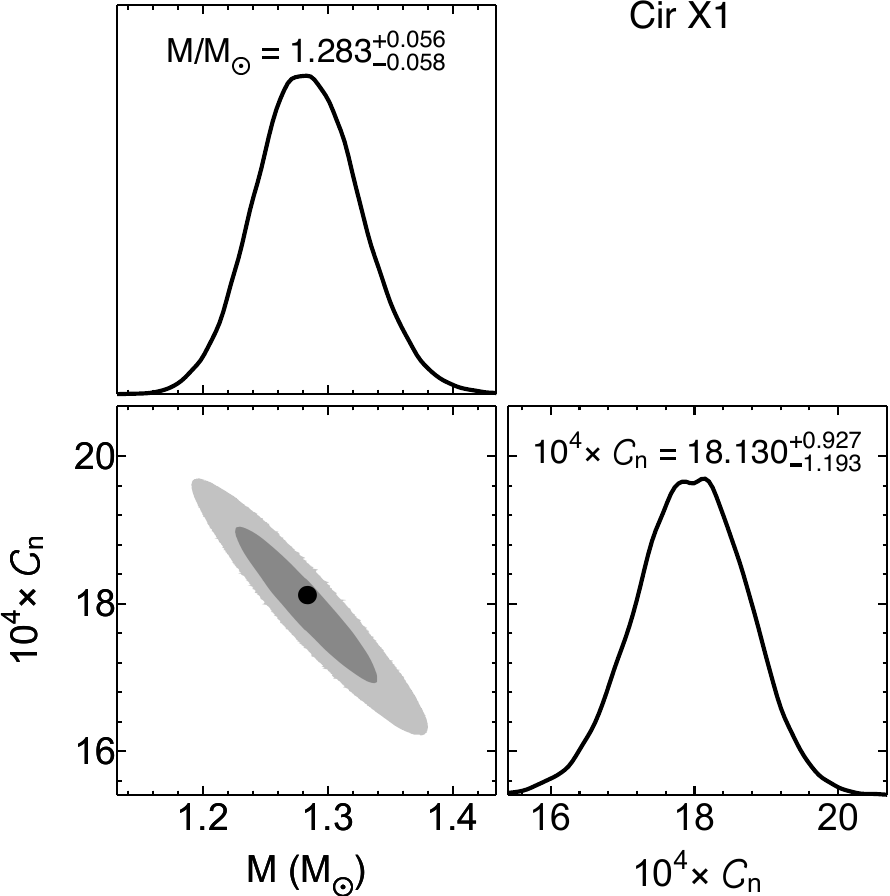}
\hfill
\includegraphics[width=0.325\hsize,clip]{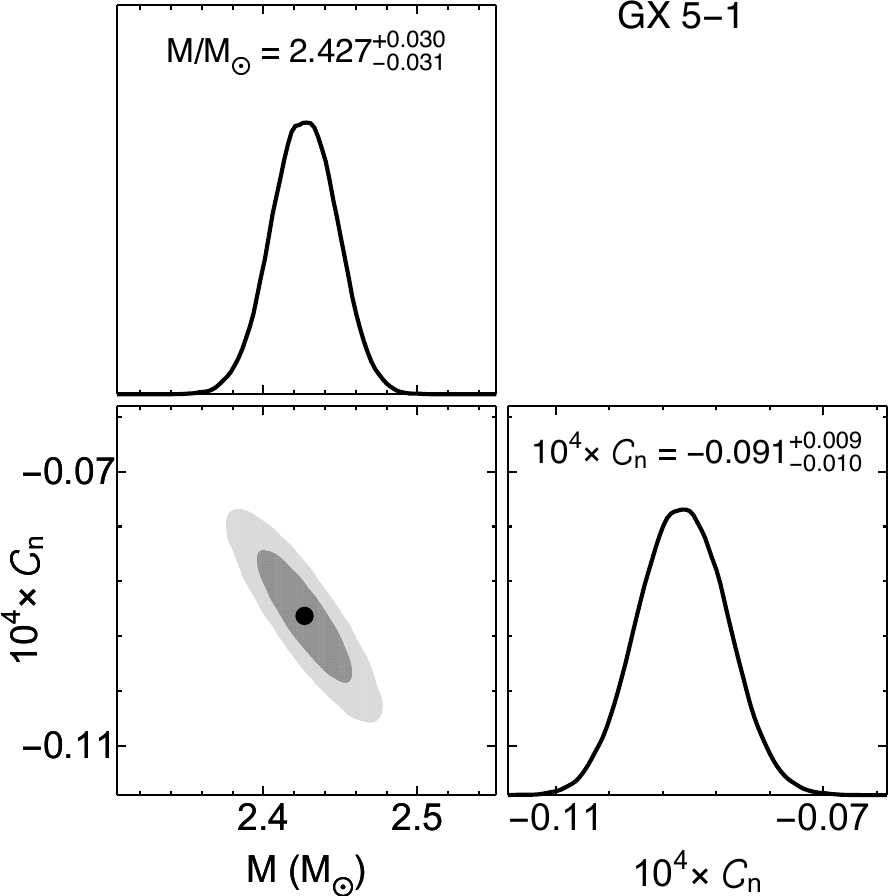}
\hfill
\includegraphics[width=0.325\hsize,clip]{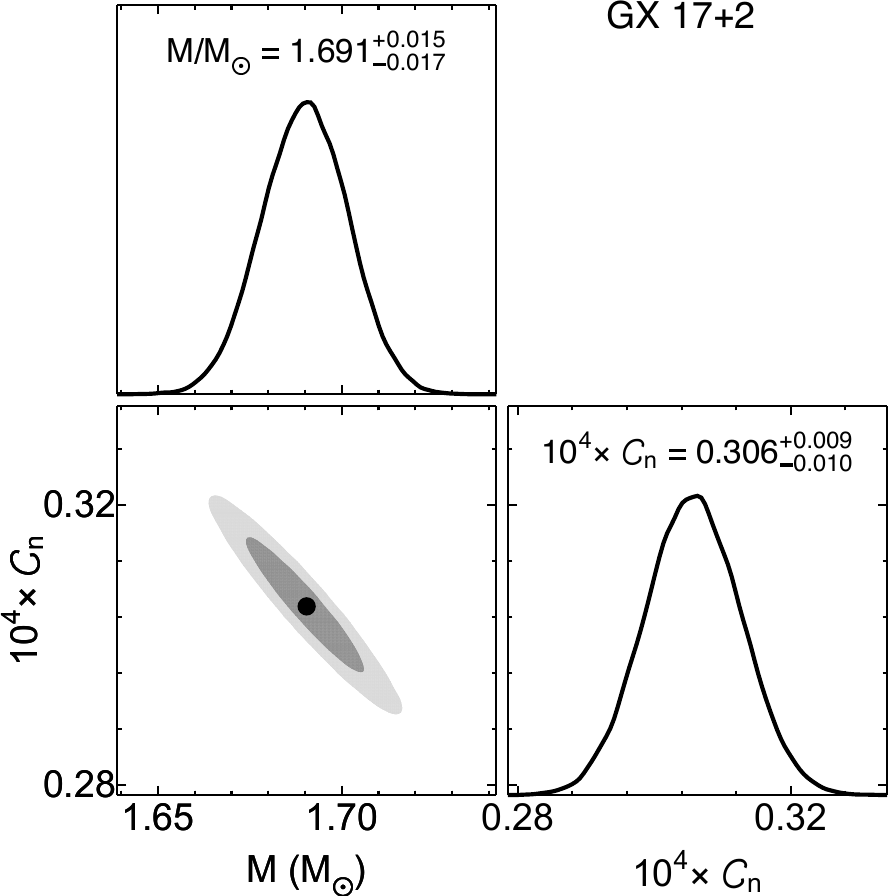}\\
\vspace{0.2cm}
\includegraphics[width=0.325\hsize,clip]{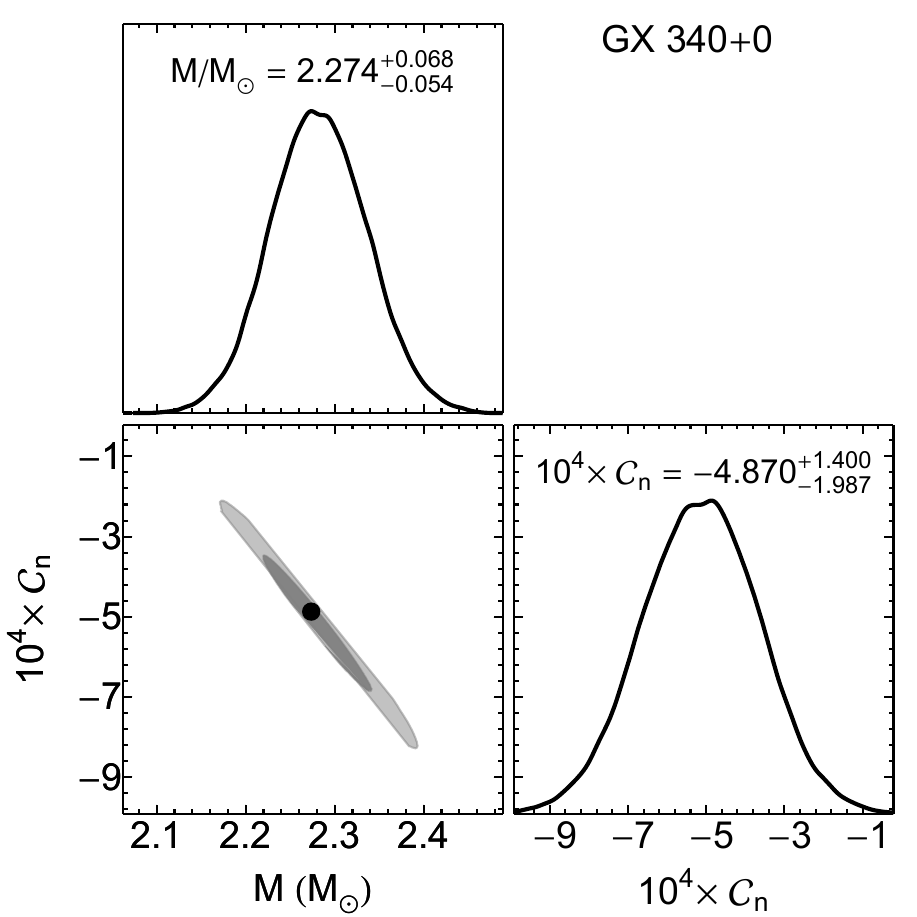}
\hfill
\includegraphics[width=0.325\hsize,clip]{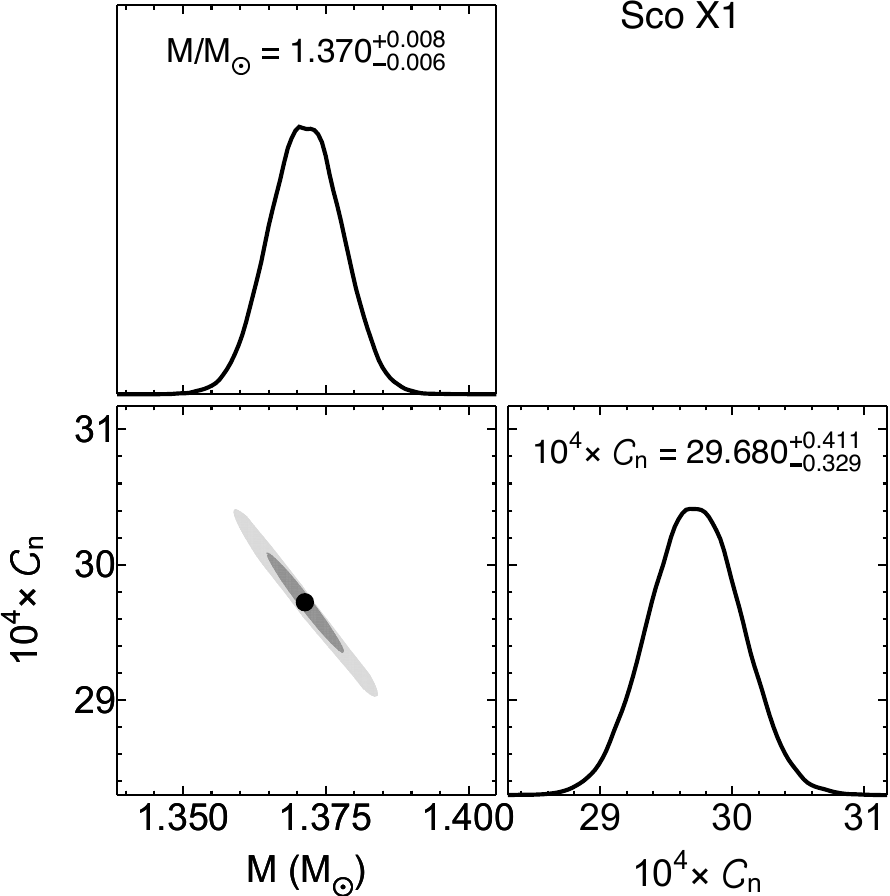}
\hfill
\includegraphics[width=0.325\hsize,clip]{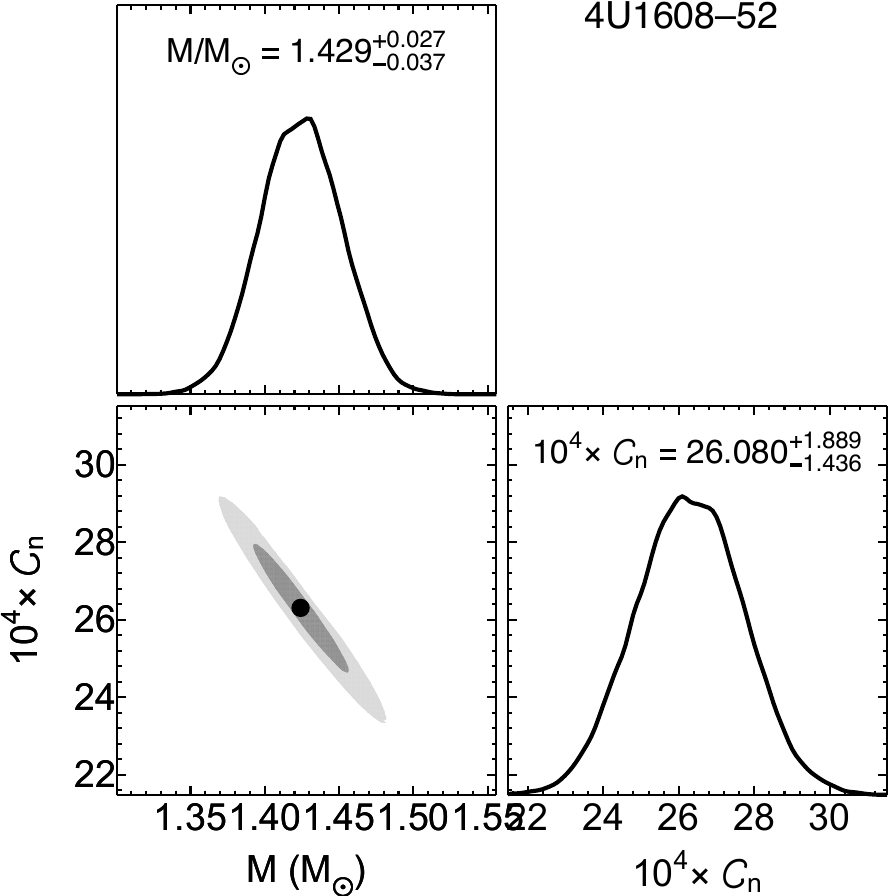}\\
\vspace{0.2cm}
{\includegraphics[width=0.325\hsize,clip]{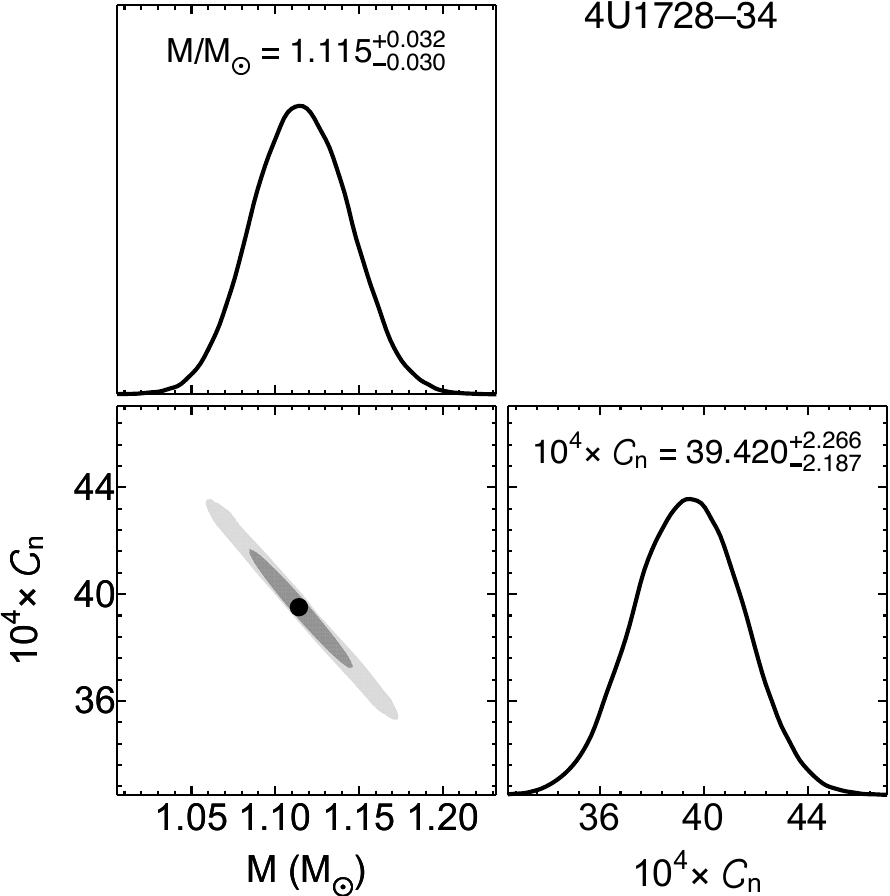}
\includegraphics[width=0.325\hsize,clip]{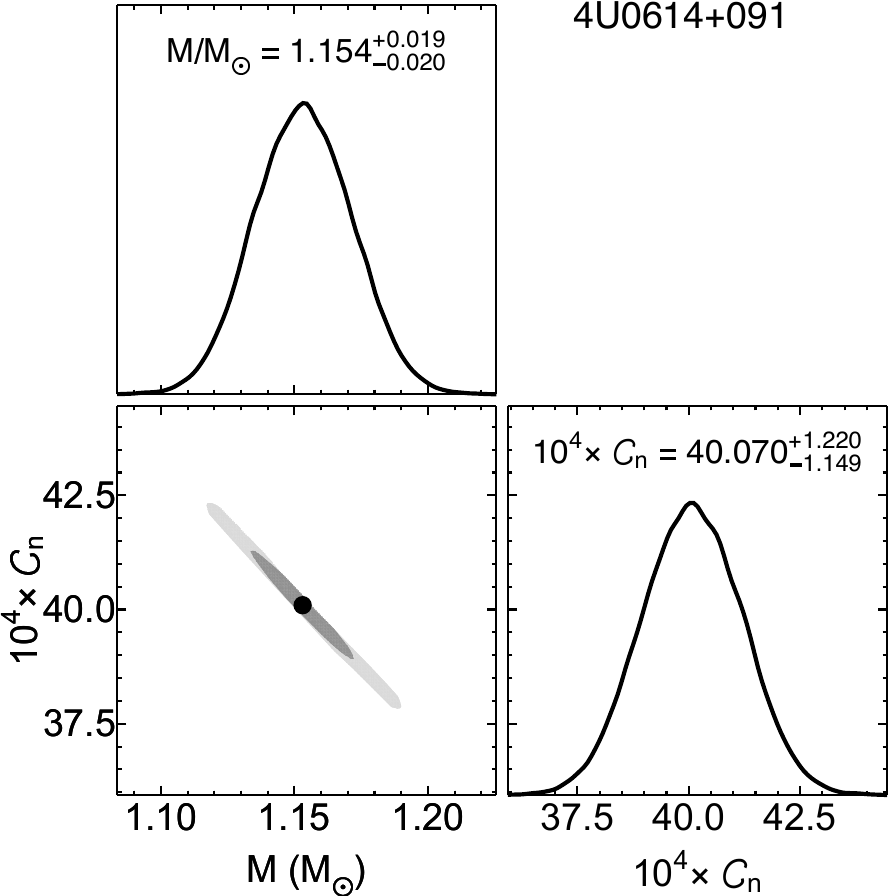}
\hfill}\\
\caption{Contour plots for each source, with best fit values, likelihoods and $2\sigma$ confidence levels. Masses are reported in solar masses, whereas the dimensions of $\mathcal C_n$ depend on the choice of $n$. }
\label{fig:MCMC}
\end{figure*}

\end{document}